\documentclass{iopjournal}
\usepackage{amsmath}
\usepackage{mathtools}
\usepackage{amssymb}
\usepackage{mathrsfs}
\usepackage{amsthm}
\usepackage{amsfonts}
\usepackage{cases}
\usepackage{bm}
\usepackage{bbm}

\usepackage{graphics}
\usepackage{graphicx}
\usepackage{graphicx,floatrow,subfig}
\usepackage{picinpar}
\usepackage{subfig}

\usepackage{physics}

\usepackage{ragged2e}
\usepackage{hyphenat}
\AtBeginDocument{
    \justifying
    \sloppy
    \hyphenpenalty=1000
    \tolerance=1000
    \emergencystretch=3em
    \renewenvironment{abstract}{%
        \vspace{16pt plus3pt minus3pt}
        {\color{gray}\hrule} \ \\
        \noindent \fontsize{11}{12}\selectfont {\bfseries Abstract}\\
        \rm\ignorespaces \justifying\sloppy\hyphenpenalty=1000\tolerance=1000\emergencystretch=3em}{\vspace{3mm} {\color{gray}\hrule}}
}

\begin{document}

\articletype{Paper}

\title{A discussion on the symmetry of relativistic Vlasov gas and its accretion in Kerr-Newman black hole}

\author{Yongqiang Liu$^1$\orcid{0009-0003-4771-0560}}

\affil{$^1$Department of Basic Courses, Wuhan Donghu University, 301 Wenhua Avenue, Wuhan City, Hubei Province 430212, China}

\email{liuyq8128@outlook.com}


\begin{abstract}
We investigate the kinetic properties of collisionless Vlasov gas in Kerr-Newman spacetime, analyzing how spacetime symmetries constrain the distribution functions. The distribution function is shown to depend on the constants of motion ($m, E, L_z, L$) and a configuration variable $Q^{3}$. Furthermore, we discuss in detail the different scenarios of charged particles being captured or scattered by the black hole in Kerr-Newman spacetime, and analytically derive the boundaries of the absorption and scattering domains in phase space. Within the Locally Non-Rotating Frame, we compute particle number density, energy density, principal pressures, and construct a set of diagnostic parameters to quantitatively measure the anisotropy information of the Vlasov gas (relative to a perfect fluid), and provide the analytic results of these physical quantities at spatial infinity. In addition, we discuss the accretion rates of the Kerr-Newman black hole and obtain their analytic expressions. Numerical results for the relative (normalized) mass and energy accretion rates reveal an identical parametric dependence: both are suppressed as the black hole's rotation $a$ and charge $Q$ increase. Conversely, the absolute value of the normalized angular momentum accretion rate (which is negative) increases with $a$, and $Q$ also moderately influences the angular momentum accretion rate through its effect on spacetime geometry. Accretion of weakly charged plasma drives charged black holes toward electrical neutrality while reducing angular momentum, ultimately favoring evolution toward Schwarzschild configurations. These findings provide new insights into kinetic accretion processes in spacetime geometries.
\end{abstract}

\section{Introduction}

	Research on accretion can be traced back to the early work of Lyttleton, Hoyle and Bondi, who first provided a relativistic solution for perfect fluids \cite{Hoyle(1939), Bondi(1944), Bondi(1952)}, and these studies were later extended to general relativity by Michel \cite{Michel(1972)}. When particle collisions become infrequent, the perfect fluid approximation breaks down. In such collisionless regimes, kinetic theory governed by the Vlasov equation provides a physically viable alternative.
	
		The theory of gas dynamics was first introduced into special relativity by J\"uttner \cite{Juttner(1911)},  and was later extended by Synge, Tauber, Israel, and others, gradually incorporating into general relativity \cite{Synge(1934), Tauber(1961), Israel(1963), Ehlers(1971), Ehlers(1973), Cercignani(2002)}. The study of the kinetic theory of gases helps to understand the properties of accretion disks and the jet launching mechanism, such as reconstructing images of the supermassive black holes at the center of the M87 galaxy \cite{Narayan(1995)}.  Furthermore, observations indicate that the gas surrounding M87 and Sgr A* is nearly collisionless and magnetized \cite{Akiyama(2019), Akiyama(2022)}.
For studies on collisionless gas models in galaxies and galaxy clusters, see Refs. \cite{Rein(2014),Hadvzic(2021)}.

	Relativistic Vlasov gases have been extensively investigated in recent years. A foundational paradigm was established by Sarbach and collaborators, who first analyzed relativistic kinetic gases and their distribution functions, and then constructed a rigorous geometric framework on the tangent and cotangent bundles \cite{Sarbach(201303), Sarbach(201309), Sarbach(201311), Sarbach(2022)}. Rioseco implemented the framework numerically in the Schwarzschild spacetime, revealing stark contrasts between Vlasov gas accretion and perfect fluid accretion. These results partially explain the observed accretion rates substantially below the Bondi predictions \cite{Rioseco-Sarbach(20171), Rioseco-Sarbach(20172)}. Gamboa \cite{Gamboa-Sarbach(2021)} and Liao \cite{Liao(2022)} incorporated angular momentum distributions, addressing phenomena such as low luminosity. Mach et al. further modeled accretion onto moving Schwarzschild black holes using Vlasov gases for dark matter simulations, achieving robust results via Monte Carlo methods \cite{Mach(20211), Mach(20212), Mach(2022), Mach(2023), Mach(2024)}.  

	For non‑Schwarzschild black holes, Cie{\'s}lik  \cite{Cieslik-Mach(2020)} extended the Vlasov accretion model to the Reissner‑Nordstr\"om spacetime, assuming a neutral J\"uttner‑distributed gas, and showed that the black hole charge suppresses the mass accretion rate. Li et al. \cite{Li(2025)} further considered a single‑species charged Fermi gas in Reissner‑Nordstr\"om spacetime, introducing the concept of a charge accretion rate. However, both works are restricted to spherical symmetry and do not include the effect of a neutral plasma (i.e., two species of opposite charges) or the rotation of the black hole.
In stationary axisymmetric spacetimes, they developed Kerr accretion models for gas confined to the equatorial plane \cite{Cieslik-Mach(2022), Mach-Odrzywolek(2023)} and unbound configurations \cite{Li(2023)}. Discussions regarding the phase space structure of gas in stationary axisymmetric spacetimes and related applications can be found in Refs. \cite{Rein(2014),Rioseco-Sarbach(2023),Rein(2023),Amesl(2016),Cai(2023)}.
Mach et al. investigated the accretion of neutral gas in Kerr spacetime and computed the full accretion rates, specifically treating slowly rotating Kerr black holes as well as analyzing Bondi-type accretion in the Kerr spacetime background \cite{Mach(20261), Mach(20262)}.

	We study a new model: a Kerr-Newman black hole accreting a collisionless plasma from infinity. The gas is assumed sufficiently dilute and the particle charges sufficiently small that particles can be treated as test particles, with gravity dominating over electromagnetic forces at infinity. We find that, 
although a single-species charged gas deviates from neutral-gas behavior due to electromagnetic interactions, in a neutral plasma the deviations of the positively and negatively charged components cancel exactly to leading order. Consequently, the plasma's macroscopic dynamics are, to leading order, identical to those of a neutral gas, while a residual net charge current slowly neutralizes the black hole charge.
 	
	This paper is organized as follows. Sec.\,\ref{Sec2} develops the theoretical framework, demonstrating how spacetime symmetries constrain the distribution function in Kerr-Newman spacetime. Sec.\,\ref{Sec3} analyzes charged particle trajectories and phase space structure. Sec.\,\ref{Sec4} derives physical observables including particle number density and pressures, discusses the diagnostic parameters for the deviation of the Vlasov gas from a perfect fluid, and finally presents the detailed expressions for the accretion rates. Sec.\,\ref{Sec5} presents numerical results profiling these quantities and their implications for black hole evolution. Finally, Sec.\,\ref{Sec6} offers a concluding summary of the work.

		We adopt these conventions:  A 4-dimensional  Lorentzian manifold is denoted by $(M,g)$, with local coordinates $x^{\mu}$ (where Greek letters $\mu,\nu,...\in \{0,1,2,3\}$). The metric signature is $(-,+,+,+)$, with geometric units $G=c=1$. The tangent bundle is $TM$, with $T_{x}M$ denoting the tangent space at $x\in M$. The cotangent bundle is $T^*M$, with $T^*_xM$ denoting the cotangent space at $x$. Lie derivatives along a vector field $v$ are denoted $\mathscr{L}_v$. Lowercase (uppercase) Latin  indices are abstract indices for tensors on $M$ ( $T^{*}M$).

\section{Kinetic theory of Vlasov gas in Kerr-Newman spacetime}\label{Sec2}

\subsection{The Vlasov equation and spacetime symmetries}

The dynamics of a collisionless gas in curved spacetime is governed by the Vlasov equation on phase space. In a spacetime $(M,g)$, the dynamical state of a particle with mass $m$ is characterized by its position $x\in M$ and physical momentum $p_a \in T^{*}_{x}M$. The physically admissible states constitute the future mass shell:
\begin{align}
\Gamma^{+}_{m}\coloneqq \left\{(x,p_{a})\in T^{*}M  \mid x\in M, \ p_{a}\in F^{+}_{x}(m)\right\},
\end{align}
where $F^{+}_{x}(m) \coloneqq \left\{p\in T^{*}_{x}M \mid  g^{ab}p_{a}p_{b}=-m^{2},\, p^{a}\equiv g^{ab}p_{b} \text{ is future-directed} \right\}$ denotes the future mass hyperboloid.

The one-particle distribution function is a mapping:
\begin{align}
f \colon \Gamma^{+}\to \mathbb{R}, \qquad  \Gamma^{+}\equiv \bigcup_{m>0} \Gamma^{+}_{m}.
\end{align}

For charged particles with mass $m$ and charge $q$ in Kerr-Newman spacetime with electromagnetic field $F_{ab} = (\text{d}A)_{ab}$, the equations of motion are:
\begin{align}
T^{c}\nabla_{c}T^{a}=\frac{q}{m} F^{a}{}_{ b}T^{b}, \label{mq}
\end{align}
where $T \equiv \frac{\partial \ }{\partial \tau}$ is the tangent vector with affine parameter $\tau$, $F^{a}{}_{b}=g^{ab}F_{ab}$. The distribution function $f$, being a one-particle distribution function for collisionless gas, remains constant along individual particle worldlines. This fundamental property is expressed by the Vlasov equation:
\begin{align}
\mathscr{L}_{\hat{T}}f=\frac{\text{d}f\circ \hat{C}(\tau)}{\text{d}\tau}=0,  \label{Vlasov}
\end{align}
where $\hat{T}$ denotes the tangent vector field to the lifted curve $\hat{C}(\tau)$ in $T^{*}M$.

To derive the explicit form, we compute the components of $\hat{T}$. The trajectory $C(\tau) \equiv \left\{ x^{\mu}(\tau) \right\}$ induces the lifted curve on $T^{*}M$:
\begin{align}
\hat{C}(\tau) \coloneqq \left\{ (x^{\mu}(\tau), p_{\nu}(\tau)) \mid p_{\mu}(\tau)\equiv m g_{\mu \nu}|_{x(\tau)}T^{\nu}(\tau) \right\},
\end{align}
where $T^{\mu} \equiv \frac{\text{d}x^{\mu}}{\text{d}\tau}$. The tangent vector field is:
\begin{align}
\hat{T}^{A}=\left( \frac{\partial \ }{\partial \tau} \right)^{A}&=\frac{\text{d}x^{\mu}}{\text{d}\tau} \left( \frac{\partial \ }{\partial x^{\mu}} \right)^{A}+\frac{\text{d}p_{\nu}}{\text{d}\tau} \left( \frac{\partial \ }{\partial p_{\nu}} \right)^{A}\notag\\
 &=g^{\mu \sigma}p_{\sigma} \left( \frac{\partial \ }{\partial x^{\mu}} \right)^{A}-\left[ \frac{1}{2}p_{\alpha}p_{\beta}\frac{\partial g^{\alpha \beta} }{\partial x^{\nu}}+qp_{\gamma}F^{\gamma}{}_{\nu} \right]\left( \frac{\partial \ }{\partial p_{\nu}} \right)^{A}.
\end{align}
The momentum evolution follows from the equations of motion:
\begin{align}
m \frac{\text{d}p_{\nu}(\tau)}{\text{d}\tau} =\frac{1}{2}p^{\rho}p^{\sigma}\partial_{\nu}g_{\rho \sigma}+qF_{\nu \gamma}p^{\gamma}=-\frac{1}{2}p_{\alpha}p_{\beta}\frac{\partial g^{\alpha \beta} }{\partial x^{\nu}}-qp_{\gamma}F^{\gamma}{}_{ \nu}.
\end{align}

This yields the explicit form of the Vlasov equation:
\begin{align}
0=m \mathscr{L}_{\hat{T}}f &=g^{\mu \nu}p_{\nu}\frac{\partial f }{\partial x^{\mu}}-\frac{1}{2}p_{\alpha}p_{\beta}\frac{\partial g^{\alpha \beta} }{\partial x^{\mu}}\frac{\partial f }{\partial p_{\mu}}-qp_{\rho} F^{\rho}{}_{ \sigma}\frac{\partial f }{\partial p_{\sigma}}. \label{Vlasov-Maxwell}
\end{align}

	The Kerr-Newman spacetime admits Killing vector fields generating isometries $\phi_k$. Crucially, under these isometries, a particle worldline $C(\tau)$ transforms to another valid worldline $\phi_k \circ C(\tau)$ due to the invariance of the equations of motion. This identification of trajectories under spacetime symmetries implies that the distribution function must satisfy the symmetry constraint:
\begin{align}
\mathscr{L}_{\hat{K}}f=0,  \label{symmetry}
\end{align}
where $\hat{K}$ is the lift of the Killing vector field $K$ to the cotangent bundle. The explicit form of $\hat{K}$ is derived as follows. 

	The coordinates transform as:
\begin{align}
(\tilde{x}^{\mu}, \tilde{p}_{\nu}(\tilde{x}))\equiv \left( \phi^{\mu}_{k}(x), (\phi_{k*}p)_{\nu}(\tilde{x})\right)=\left( \phi^{\mu}_{k}(x), \frac{\partial x^{\alpha} }{\partial \tilde{x}^{\nu}}p_{\alpha}(x)\right),
\end{align}
where $\phi_{k*}$  denotes the pushforward map induced by the isometry $\phi_{k}$.
Differentiating with respect to $k$ at $k=0$:
\begin{align}
\hat{K}^{A}=\left(\frac{\partial \ }{\partial k}\right)^{A} &=\left. \frac{\text{d}\tilde{x}^{\mu}}{\text{d}k}\right|_{k=0} \left(\frac{\partial \ }{\partial x^{\mu}}\right)^{A}+\left. \frac{\text{d}\tilde{p}_{\nu}}{\text{d}k}\right|_{k=0} \left(\frac{\partial \ }{\partial p_{\nu}}\right)^{A} \notag \\
& = K^{\mu} \left(\frac{\partial \ }{\partial x^{\mu}}\right)^{A}-p_{\alpha}\frac{\partial K^{\alpha} }{\partial x^{\nu}} \left(\frac{\partial \ }{\partial p_{\nu}}\right)^{A}. \label{hatK}
\end{align}

The first term follows directly: $\left. \frac{\text{d}\tilde{x}^{\mu}}{\text{d}k}\right|_{k=0}=\left. K^{\mu}(\tilde{x})\right|_{k=0}=K^{\mu}(x)$. For the second term, from $x=\phi_{-k}(\tilde{x})$, we compute:
\begin{align*}
\frac{\text{d}\ }{\text{d}k}\frac{\partial x^{\alpha} }{\partial \tilde{x}^{\nu}}=\frac{\text{d}\ }{\text{d}k}\frac{\partial \phi^{\alpha}_{-k}(\tilde{x}) }{\partial \tilde{x}^{\nu}}=\frac{\partial \ }{\partial \tilde{x}^{\nu}}\frac{\text{d}\phi^{\alpha}_{-k}(\tilde{x})}{\text{d}k} =-\frac{\partial \ }{\partial \tilde{x}^{\nu}}K^{\alpha}(\phi_{-k}(\tilde{x}))=-\frac{\partial x^{\beta} }{\partial \tilde{x}^{\nu}}\frac{\partial K^{\alpha}(x) }{\partial x^{\beta}},
\end{align*}
since $\phi_{0}=\text{id}$ is the identity map. Therefore:
\begin{align}
\left. \frac{\text{d}\tilde{p}_{\nu}}{\text{d}k}\right|_{k=0}=p_{\alpha} \frac{\partial K^{\alpha} }{\partial x^{\nu}},
\end{align}
which completes the derivation of Eq.\eqref{hatK}.

The constraint on the distribution function from spacetime symmetry becomes:
\begin{align}
0=\mathscr{L}_{\hat{K}}f=K^{\mu}\frac{\partial f }{\partial x^{\mu}}-p_{\alpha} \frac{\partial K^{\alpha} }{\partial x^{\nu}} \frac{\partial f }{\partial p_{\nu}}. \label{Killing}
\end{align}

\subsection{Constants of motion and integrability}

The Kerr-Newman metric in Boyer-Lindquist coordinates $\{t,r,\theta,\varphi\}$ is:
\begin{align}
\text{d}s^{2}=-\frac{\varDelta-a^{2}\sin^{2} \theta}{\rho^{2}}\text{d}t^{2}-\frac{2a\sin^{2}\theta (r^{2}+a^{2}-\varDelta)}{\rho^{2}}\text{d}t \text{d}\varphi +\frac{\varSigma^{2}}{\rho^{2}} \sin^{2}\theta \text{d}\varphi^{2}+\frac{\rho^{2}}{\varDelta }\text{d}r^{2}+\rho^{2} \text{d}\theta^{2},\label{Kerr}
\end{align}
where $\rho^{2}  = r^{2}+a^{2}\cos^{2} \theta$, $\varDelta = r^{2}-2M r+a^{2}+Q^{2}$, and $\varSigma^{2} = (r^{2}+a^{2})^{2}- \varDelta a^{2} \sin^{2}\theta$.  The electromagnetic field compatible with the Kerr-Newman spacetime is given by $A_{b}=-\frac{Qr}{\rho^{2}}\bqty{(\dd{r})_{b}-a\sin^{2}\theta (\dd{\varphi})_{b} }.$

The spacetime symmetries provide four independent constants of motion:
\begin{align}
E &\coloneqq -\pi_{a}\xi^{a} , & L_{z} &\coloneqq \pi_{a}\psi^{a}, & H &\coloneqq \frac{1}{2}g^{ab}p_{a}p_{b}, &L^{2}  &\coloneqq C^{ab}p_{a}p_{b}, 	 \label{constants}
\end{align}
where $\xi^a = (\partial_t)^a$ and $\psi^a = (\partial_\varphi)^a$ are the Killing vectors associated with stationarity and axisymmetry, $\pi_{a}=p_{a}+q A_{a}$ is the canonical momentum, and $C^{ab}$ is the Killing tensor responsible for the hidden symmetry\footnote{In Boyer--Lindquist coordinates, an explicit calculation gives $C_{ab}F^{b}{}_{c}=-C_{cb}F^{b}{}_{a}$.} \cite{Carter(1973)}:
\begin{align}
C^{ab}\equiv 2\rho^{2} l^{(a}k^{b)}+r^{2}g^{ab},
\end{align}
with the principal null directions:
\begin{align}
l^{a} &\equiv \frac{1}{\varDelta}[(r^{2}+a^{2})\xi^{a}+a \psi^{a}+\varDelta (\frac{\partial \ }{\partial r})^{a}], &
k^{a} &\equiv \frac{1}{2\rho^{2}}[(r^{2}+a^{2})\xi^{a}+a \psi^{a} -\varDelta (\frac{\partial \ }{\partial r} )^{a}].
\end{align}

To understand how these symmetries constrain the distribution function, we examine the phase space structure. The symplectic form on phase space is:
\begin{align}
\Omega_{AB}\coloneqq [\text{d}(\pi_{\mu}(\text{d}x^{\mu}))]_{AB}=(\text{d}\pi_{\mu})_{A}\wedge (\text{d}x^{\mu})_{B}=(\text{d}p_{\mu})_{A}\wedge (\text{d}x^{\mu})_{B}+q \hat{F}_{AB},
\end{align}
where $\hat{F}_{AB}\equiv \frac{1}{2}F_{\mu \nu}(\text{d}x^{\mu})_{A}\wedge (\text{d}x^{\nu})_{B}$.

The Hamiltonian vector field $Z_{\mathcal{F}}$ induced by a function $\mathcal{F}(x,\pi)$ is given by:
\begin{align}
(\text{d}\mathcal{F})_{B}&= -Z^{A}_{\mathcal{F}}\Omega_{AB}, & Z^{A}_{\mathcal{F}}&=\Omega^{AB}(\text{d}\mathcal{F})_{B}.
\end{align}
The dual of the symplectic form $\Omega^{AB}$ is defined by
\begin{align}
\Omega^{AB}\Omega_{BC}&=\delta^{A}{}_{C}, & \Omega^{AB}&=\left(\frac{\partial \ }{\partial x^{\nu}}\right)^{A}\wedge \left(\frac{\partial \ }{\partial \pi_{\mu}}\right)^{B}=\left(\frac{\partial \ }{\partial x^{\nu}}\right)^{A}\wedge \left(\frac{\partial \ }{\partial p_{\mu}}\right)^{B}+q\hat{F}^{AB},
\end{align}
where $\hat{F}^{AB}\equiv \frac{1}{2}F_{\mu \nu}\left(\frac{\partial \ }{\partial p_{\mu}}\right)^{A}\wedge \left(\frac{\partial \ }{\partial p_{\nu}}\right)^{B}$.

Direct computation gives the Hamiltonian vector fields for the constants of motion:
\begin{align}
Z_{\pi(K)}^{A}&=K^{\mu} \left(\frac{\partial \ }{\partial x^{\mu}}\right)^{A}-p_{\alpha}\frac{\partial K^{\alpha} }{\partial x^{\nu}}\left(\frac{\partial \ }{\partial p_{\nu}}\right)^{A}=\hat{K}^{A},\\
Z^{A}_{K(p,p)}&=p_{\alpha}K^{\alpha \mu}\left(\frac{\partial \ }{\partial x^{\mu}}\right)^{A}-\frac{1}{2}p_{\alpha}p_{\beta}\frac{\partial K^{\alpha \beta} }{\partial x^{\nu}}\left(\frac{\partial \ }{\partial p_{\nu}}\right)^{A}-qp_{\alpha}K^{\alpha \beta}F_{\beta \nu}\left(\frac{\partial \ }{\partial p_{\nu}}\right)^{A}.
\end{align}

The Vlasov equation is $\mathscr{L}_{Z_{H}}f=0$, and the Killing constraints are $\mathscr{L}_{Z_{E}}f=0$ and $\mathscr{L}_{Z_{L_{z}}}f=0$. Although $L^{2}$ is also a second-order Killing invariant, it does not imply $\mathscr{L}_{Z_{L^{2}}}f=0$; rather, it only guarantees that $\mathscr{L}_{Z_{L^{2}}}f$ is a constant of motion, i.e.,
\begin{align}
\mathscr{L}_{Z_{H}}\pqty{\mathscr{L}_{Z_{L^{2}}}f}=\mathscr{L}_{Z_{H}}\pqty{\mathscr{L}_{Z_{L^{2}}}f}-\mathscr{L}_{Z_{L^{2}} }\pqty{\mathscr{L}_{Z_{H} }f}=\mathscr{L}_{[Z_{H}, Z_{L^{2}}]}f=\pqty{\mathscr{L}_{Z_{H}}	Z_{L^{2}}}(f)=0,
\end{align}
a conclusion that will later be seen as rather trivial. The proof of the last step is as follows:
Since the symplectic form $\Omega$ is non-degenerate, we prove $\Omega_{AB}\mathscr{L}_{Z_{H}}Z^{A}_{L^{2}}=0$:
\begin{align}
\Omega_{AB}\left( \mathscr{L}_{Z_{H}}Z_{L^{2}}\right)^{A} &=\mathscr{L}_{Z_{H}}\left(Z^{A}_{L^{2}}\Omega_{AB}\right)-Z^{A}_{L^{2}}\left( \mathscr{L}_{Z_{H}} \Omega\right)_{AB} \notag \\
&=\mathscr{L}_{Z_{H}}(-\text{d}_{B}L^{2})-Z^{A}_{L^{2}}\left(\text{d}_{A}(Z^{C}_{H}\Omega_{CB})+Z^{C}_{H}(\text{d}\Omega)_{CAB} \right) \notag \\
&=-\text{d}_{B}(\mathscr{L}_{Z_{H}}L^{2})-Z^{A}_{L^{2}}(-\text{d}_{A}\text{d}_{B}H)=0.
\end{align}

\subsection{Canonical coordinates and general form of distribution function}

	The complete integrability of the system allows us to construct action-angle variables. Defining canonical momentum coordinates:
\begin{align}
P_{0} &\equiv \sqrt{-2H},& P_{1}&\equiv E,& P_{2}&\equiv L_{z}, &	P_{3}&\equiv L=\sqrt{L^{2}}, \label{canP}
\end{align}
and their conjugate configuration variables $Q^\mu$ via the action functional $S  = \frac{1}{2}m^{2}\tau + \int \pi_{\mu}(x^{\alpha}, P_{\beta})\text{d}x^{\mu}$ \cite{Straumann(2013)}, $Q^{\mu}\coloneqq \frac{\partial S }{\partial P_{\mu}}$, the symplectic form simplifies to:
\begin{align}
\varOmega_{AB}=\left(\frac{\partial \pi_{\alpha} }{\partial P_{\mu}}\right)\text{d}P_{\mu}\wedge \text{d}x^{\alpha}=\frac{\partial ^{2}S\  }{\partial  P_{\mu}\partial x^{\alpha} }\text{d}P_{\mu}\wedge \text{d}x^{\alpha}=\frac{\partial Q^{\mu} }{\partial x^{\alpha}}\text{d}P_{\mu}\wedge \text{d}x^{\alpha}=\text{d}P_{\mu}\wedge \text{d}Q^{\mu}.
\end{align}
In these canonical coordinates, the phase space dynamics simplifies considerably: the momenta $P_\mu$ (constants of motion) remain fixed along a particle's phase space trajectory, while the conjugate coordinates $Q^\mu$ evolve linearly with the affine parameter. This structure explicitly manifests the complete integrability of the system, even in the presence of electromagnetic interactions.

	In these coordinates, the Hamiltonian vector fields become:
\begin{align}
Z_{H}&=P_{0}\frac{\partial \ }{\partial Q^{0}}, & Z_{E}&=-\frac{\partial \ }{\partial Q^{1}}, & Z_{L_{z}}&=\frac{\partial \ }{\partial Q^{2}}, & Z_{L^{2}}&=2P_{3} \frac{\partial \ }{\partial Q^{3}}.
\end{align}
The Vlasov equation $\mathscr{L}_{Z_H}f=0$ and the Killing constraints $\mathscr{L}_{Z_E}f=0$, $\mathscr{L}_{Z_{L_z}}f=0$ imply:
\begin{align}
 \frac{\partial f }{\partial Q^{0}} &=0,&  \frac{\partial f }{\partial Q^{1}}&=0, & \frac{\partial f }{\partial Q^{2}}&=0.
\end{align}
In our technique of excluding configuration variables, the earlier conclusion that $\mathscr{L}_{Z_{L^{2}}}f$ is a constant of motion cannot be directly used to judge whether $f$ depends on the configuration variable $Q^{3}$, because
\begin{align}
\mathscr{L}_{Z_{H}} Z_{L^{2}}(f) =P_{0}\pdv{Q^{0}}\pqty{ 2P_{3} \pdv{Q^{3}}f}&=2P_{0}P_{3}\frac{\partial ^{2}f }{\partial Q^{0} \partial Q^{3}} \notag\\
&=2P_{0}P_{3} \pdv{Q^{3}}\pqty{\pdv{Q^{0}}f}=2P_{0}P_{3} \pdv{Q^{3}}0=0.
\end{align}
Thus we obtain the functional form of the distribution function
\begin{align}
f=f( P_{0}, P_{1}, P_{2}, P_{3}; Q^{3}). \label{fQPPPP}
\end{align}
Because the Vlasov gas is collisionless, the distribution function at a finite location does not correspond to any local thermodynamic equilibrium; it is a number density function describing the state of particles in phase space. At infinity, one may discuss a thermodynamic equilibrium distribution for the particle source, and this is then extended to finite locations via the Vlasov equation, yielding the distribution function that we analyze.

	In the above derivation, we not only assumed that the charged particles are test particles in the Kerr-Newman spacetime, but also neglected the electromagnetic field generated by the gas itself; consequently, \eqref{fQPPPP} is not a solution of the full Einstein-Maxwell-Vlasov equations \cite{Sarbach(201311), Sarbach(2022)}. As long as the gas is sufficiently dilute and the particle charges are sufficiently small, our simplified treatment is reliable.

In the Schwarzschild spacetime, the additional Killing vector fields allow one to further exclude the variables $Q^{3}$ and $P_{2}$; see \cite{Rioseco-Sarbach(20171)} for details. Even for a Schwarzschild black hole, the situation changes when the black hole moves relative to the background. This case was discussed in \cite{Mach(20211)}, where $Q^{3}$ was introduced as a configuration variable (angle variable) into the distribution function and analyzed in detail.

In what follows, we restrict our attention to distribution functions that do not depend on $Q^{3}$. In particular, we adopt the specific J\"uttner distribution and the monoenergetic distribution to address the physical problem of interest.

\section{Particle Trajectory Analysis} \label{Sec3}
	
The general form of the distribution function, $f=f( m, E, L_{z}, L; Q^{3})$, derived in the previous section, is fundamentally constrained by the spacetime symmetries of the Kerr-Newman geometry. However, to compute physical observables such as the particle current and energy-momentum tensor—which involve momentum-space integrals of $f$—we must precisely define the domain of integration at each spacetime point $x=(t_{x}, r_{x}, \theta_{x}, \varphi_{x})$. This domain is intrinsically determined by the dynamics of individual particle trajectories.

A central task in this analysis is to distinguish between two distinct classes of orbits: those corresponding to particles that will inevitably fall into the black hole (absorbed trajectories), and those that reach a periastron (radial turning point) before being scattered back to infinity (scattered trajectories). This critical demarcation is governed by the conditions for turning point, which separate the phase space into absorption and scattering regions. We therefore proceed with a detailed analysis of the equations of motion for charged particles in the Kerr-Newman spacetime.

The equations of motion for charged particles in Kerr-Newman spacetime are given by:
\begin{align}
(\rho^{2}\dot{r})^{2}&=\left[(r^{2}+a^{2}) E-a L_{z}-\kappa r \right]^{2}-\varDelta (L^{2}+m^{2}r^{2}), \label{dotr}\\
(\rho^{2}\dot{\theta})^{2}&=L^{2}-m^{2} a^{2}\cos^{2}\theta-\left( \frac{L_{z}}{\sin \theta}-a E\sin \theta\right)^{2}, \label{dottheta}\\
\rho^{2} \dot{t} &=\left(r^{2}+a^{2}\right)\cdot  \frac{(r^{2}+a^{2})E-aL_{z}-\kappa r}{\varDelta}+a\left(L_{z}-aE\sin^{2}\theta\right), \label{dot}\\
\rho^{2} \dot{ \varphi} &= a\cdot  \frac{(r^{2}+a^{2})E-aL_{z}-\kappa r}{\varDelta}+\sin^{-2}\theta \left(L_{z}-aE\sin^{2}\theta\right).\label{dotvarphi}
\end{align}
For a test particle, $\kappa \equiv \frac{qQ}{mM}$ represents the ratio of the electrostatic force to the Newtonian gravitational force at spatial infinity, $\pqty{\frac{qQ}{ r^{2}}}/\pqty{\frac{mM}{r^{2}}}$.
To facilitate our analysis, we introduce the following dimensionless parameters:
\begin{align}
a &\leftarrow \frac{a}{M}, & Q &\leftarrow \frac{Q}{M}, & x^{\mu} &\leftarrow \frac{x^{\mu}}{M}, & \tau &\leftarrow \frac{\tau}{M}, \\
E &\leftarrow \frac{E}{m}, & L_{z} &\leftarrow \frac{L_{z}}{mM}, & L^{2} &\leftarrow \frac{L^{2}}{(mM)^{2}}, & q &\leftarrow \frac{q}{m}.
\end{align}

We only consider those worldlines that come from infinity and are future-directed, which requires
\begin{align}
\displaystyle\mathop{\text{lim}}_{r \to \infty}(\dot{r})^{2}=E^{2}-1 \geq 0,  \quad  \displaystyle\mathop{\text{lim}}_{ r \to \infty}\dot{t} =E >0,
\qquad\qquad \Rightarrow \quad E \geq 1.
\end{align}
We now discuss the radial equation of motion \eqref{dotr},
\begin{align}
(\rho^{2}\dot{r})^{2}= R &\equiv \bqty{(r^{2}+a^{2}) E-a L_{z}-\kappa r}^{2}-\varDelta (L^{2}+r^{2}) . \label{dotrR}
\end{align}
If the radial orbit possesses a turning point $r_{\text{turn}}$, the corresponding critical conditions are
\begin{align}
R(r_{\text{c}}) &=0, & R'(r_{\text{c}}) &=0, \label{turnpoint}
\end{align}
where $R'(r_{\text{c}}) \equiv \left. \frac{\partial R(r) }{\partial r}\right|_{r=r_{\text{c}}}$. Solving the critical turning point equations yields $E_{c}=E_{c}(r_{\text{c}}, L_{z}), L_{c}=L_{c}(r_{\text{c}}, L_{z})$. 

Suppose that the particle's worldline passes through a spacetime point $x =( t_{x}, r_{x}, \theta_{x}, \varphi_{x})$. The constants of motion $(E, L, L_{z})$ must be such that Eqs. \eqref{dotr}--\eqref{dotvarphi} are all valid, from which the angular equation of motion \eqref{dottheta} yields, based on $\theta_{x}$, the parametrization
\begin{align}
p_{\theta}=\rho^{2} \dot{\theta}&=\bar{L} \cos \sigma, & L_{z}&=a E\sin^{2}\theta_{x} +\bar{L}\sin \theta_{x} \sin \sigma , \qquad \sigma \in [0, 2\pi), \label{barL}
\end{align}
with $\bar{L}\equiv \sqrt{L^{2}-a^{2}\cos^{2}\theta_{x} }$, and the range of $\sigma$ can also be taken as any continuous closed interval of length $2\pi$. This establishes the bijection
\begin{align}
\pqty{ E, L_{z}, L ; p_{\theta}|_{\theta=\theta_{x}}}\xrightarrow{\sim} \pqty{ \sigma, E, \bar{L}; p_{\theta}|_{\theta=\theta_{x}}}.
\end{align}
Replacing the constants of motion in Eq. \eqref{dotrR} with the new set of constant parameters $(\sigma, E, \bar{L})$ (which depends on the angular coordinate $\theta_{x}$ at $x$) gives
\begin{align}
R=\left(\rho^{2}E-a\sin\theta\sin\sigma \bar{L}  -\kappa r\right)^{2}-\varDelta (\bar{L}^{2}+\rho^{2}), \qquad  \rho^{2} \equiv r^{2}+a^{2}\cos^{2}\theta.  \label{R(r)}
\end{align}
Unless the dependence on $x$ is emphasized, the subscript $x$ of the constant $\theta_{x}$ is omitted. Solving the system of critical turning point equations \eqref{turnpoint} yields
\begin{align}
E_{\text{solve}} &=\frac{\kappa r}{\rho^{2}}+\frac{1}{\rho^{2}}\sqrt{\frac{\varDelta}{B^{2}}}+\frac{a \sin\theta\sin\sigma}{\rho^{2}}\sqrt{\frac{1}{B^{2}}-\rho^{2}}, & \bar{L}_{\text{solve}}&=\sqrt{\frac{1}{B^{2}}-\rho^{2}}. \label{ELB}
\end{align}
Physically, $E_{\text{solve}}(r)$ represents the minimum energy required for a particle to be scattered at radial position $r$ (i.e., for the orbit to possess a radial turning point), while $\bar{L}_{\text{solve}}$ is the corresponding angular momentum parameter of that critical orbit. Here $B=B(a, Q, r, a\sin\theta_{x}, a\sin\sigma, \kappa )$. $E_{\text{solve}}$ and $\bar{L}_{\text{solve}}$ depend on the same set of variables as $B$, and understanding this dependence structure facilitates the analysis. However, when we focus on a particular central variable, the other secondary variables will be omitted. $B$ satisfies the equation
\begin{align}
\varphi_{a} \sqrt{1-B^{2}\rho^{2}}=B^{2}-\varphi_{\kappa}B-\varphi,  \qquad B \geq 0,\label{solB}
\end{align}
where $\varphi =\frac{2}{\rho^{2}}-\frac{r-1}{r \varDelta}, \varphi_{a}\equiv \frac{2a\sin\theta\sin\sigma}{\rho^{2}\sqrt{\varDelta}},  \varphi_{\kappa}\equiv \frac{\kappa}{r}\cdot \frac{r^{2}-a^{2}\cos^{2}\theta}{r^{2}+a^{2}\cos^{2}\theta}\cdot \frac{1}{\sqrt{\varDelta}}. $

Let $r_{\text{ph}}$ be the radius of the photon sphere ($E\to \infty, \bar{L} \to \infty$). From Eqs. \eqref{ELB}--\eqref{solB} we have
\begin{align}
 B^{2}(r_{\text{ph}}) &=0, & \varphi(r_{\text{ph}})+\varphi_{a}(r_{\text{ph}})&=0. \label{Brph}
\end{align}
The second relation shows that the electromagnetic coupling parameter $\kappa$ does not affect the location of the photon sphere, and $r_{\text{ph}}=r_{\text{ph}}(a, Q, a\sin\theta_{x}, a\sin\sigma)$. Solving $B(a, \kappa)$ analytically involves the quartic root formula; however, when $a \cdot \kappa=0$, there are simple solutions
\begin{align}
B^{2}(0,\kappa) &=\bqty{\frac{\varphi_{\kappa}}{2}+ \sqrt{\left(\frac{\varphi_{\kappa}}{2}\right)^{2}+\varphi}}^{2},&
B^{2}(a, 0)= \varphi -\frac{\rho^{2}\varphi^{2}_{a}}{2}+\varphi_{a} \sqrt{\frac{\rho^{4}}{4} \varphi^{2}_{a} -\rho^{2}\varphi +1},
\end{align}
where $r \in (r_{\text{ph}},+\infty)$.  Here the effective expression for $B(0,\kappa)$ in the weak electromagnetic coupling limit is considered; for a general analysis, see \cite{Liu(2026)}.

\begin{figure}[htbp]
\centering
\subfloat[$E_{c}-r$]{\includegraphics[width=0.48\linewidth]{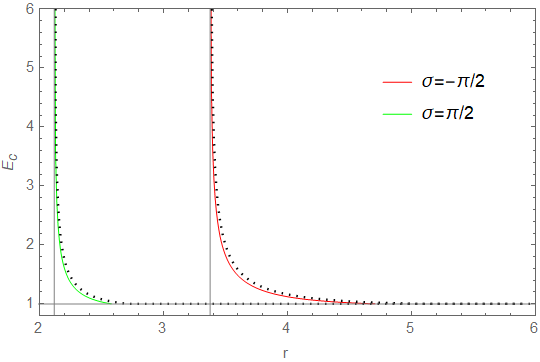} \label{figEcr}}
\hfill
\subfloat[$\bar{L}_{c}-E$]{\includegraphics[width=0.48\linewidth]{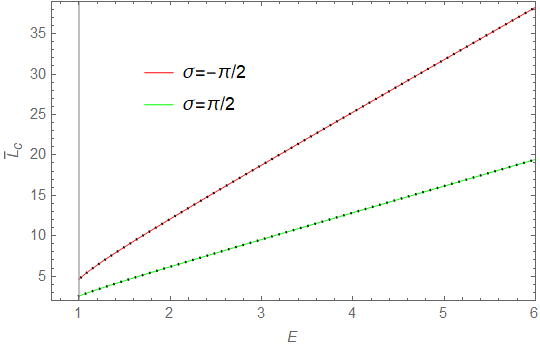} \label{figLcE}}
\caption{(a) Critical energy $E_{c}$ at the radial turning point as a function of the turning point position $r$; (b) Critical scattering parameter $\bar{L}_{c}$ in phase space as a function of energy $E$. The parameters are $a=Q=0.5, \kappa=0.1, \theta_{x}=\frac{\pi}{2},\sin \sigma=\pm 1$. Solid curves are from the numerical solution of the critical turning point equations, and black dotted curves are from the analytical formula Eq. \eqref{ELB} (accurate to order $\kappa$). The physical cutoff $E \ge 1$ is taken into account.}
\label{figELB}
\end{figure}

We consider the case where gravity dominates at infinity, i.e., $\kappa \ll 1$, in which case 
\begin{align}
B(a, \kappa)&\simeq B(a, 0)+\kappa \eval{\pdv{B(a, \kappa)}{\kappa}}_{\kappa=0}+o(\kappa), \\ \eval{\pdv{B(a, \kappa)}{\kappa}}_{\kappa=0} &\equiv \frac{1}{r\sqrt{\varDelta}} \cdot \frac{r^{2}-a^{2}\cos^{2}\theta}{r^{2}+a^{2}\cos^{2}\theta}\cdot \pqty{ 2+\frac{\rho^{2}\varphi^{2}_{a}}{B^{2}(a, 0)-\varphi}}^{-1}, 
\end{align}
where $o(\kappa)$ denotes higher-order small quantities $\Bqty{ \kappa^{2}, \kappa^{3}, \cdots}$. With $E_{\text{solve}}$ and $\bar{L}_{\text{solve}}$ accurate to order $\kappa$, their analytical curves and the exact numerical curves from the critical turning point equations are shown in Fig. \ref{figELB}.

Fig. \ref{figELB} plots the curves of relevant parameters at the turning point. The turning point refers to the turning point of scattered orbits that pass through a fixed point in the equatorial plane $(r_{x}, \theta_{x}=\frac{\pi}{2}, \varphi_{x})$ and satisfy the parameter relation $\sigma=\pm \frac{\pi}{2}$.
In Fig. \ref{figEcr}, the energy $E_{\text{solve}}$ goes to infinity at the photon sphere $r_{\text{ph}}$, $E_{\text{solve}}(r_{\text{ph}}) \to \infty$, and decreases monotonically to $E_{\text{solve}}(r_{\text{mb}})=1$ at the position $r_{\text{mb}}$. Numerically, $E_{\text{solve}}(r, \sigma, \theta, \kappa )$ decreases monotonically on the interval $(r_{\text{ph}}, r_{\text{mb}})$, and this trend is independent of the specific parameters $\sigma, \theta_{x}, \kappa$. We construct the following function
\begin{align}
E_{c}(r) &=\begin{cases}
E_{\text{solve}}(r), &  r_{\text{ph}}<r<r_{\text{mb}},\\
1, &  r \geq r_{\text{mb}},
\end{cases}&
\bar{L}_{c}(\sigma, E)&=\bar{L}_{c}(a,Q, a\sin\theta_{x}, a\sin\sigma, E,\kappa). \label{EcLc(E)}
\end{align}
For any given energy $E \ge 1$, there exists a unique radius $r_{\text{c}}(E; \kappa) \in (r_{\text{ph}}, r_{\text{mb}}]$ determined by the equation $E_{\text{solve}}(r;\kappa)=E$. From this we construct the composite function\footnote{Substituting the solutions $r_{c}(E; \kappa)$ and $\bar{L}_{c}(E;\kappa)$ into the parametric equations determined by the critical point (with $L_{z}$ parametrized by \eqref{barL}) \eqref{turnpoint}, it follows that $\bar{L}_{c}$ is differentiable with respect to $\kappa$ at $\kappa=0$. } $\bar{L}_{c}(\sigma, E; \kappa)\coloneqq\bar{L}_{\text{solve}}(r_{\text{c}}(E;\kappa); \kappa)$. Its numerical curve is shown in Fig. \ref{figLcE} and exhibits a nearly linear growth with energy $E$.

For a given spacetime point $x =( t_{x}, r_{x}, \theta_{x}, \varphi_{x})$, the radial equation of motion \eqref{R(r)} projected from the worldline determined by the parameter set $(\sigma_{0}, E_{0}, \bar{L} )$ is classified as follows:
\begin{enumerate}
\item
Critical trajectory: If $\bar{L}=\bar{L}_{c}(\sigma_{0}, E_{0}; \theta_{x})$, then solving $E_{\text{solve}}(r; \theta_{x}, \sigma_{0})=E_{0}$ gives a unique turning point $r_{\text{c}}$; this case is of measure zero and can be neglected.
\item
Captured trajectory: If $0< \bar{L} < \bar{L}_{c}(\sigma_{0}, E_{0}; \theta_{x})$, then the orbit has no radial turning point and corresponds to a particle captured by the black hole.
\item
Scattered trajectory: If $ \bar{L}_{c}(\sigma_{0}, E_{0}; \theta_{x})< \bar{L}< \bar{L}_{\max}(\sigma_{0}, E_{0}; r_{x}, \theta_{x})$, the orbit possesses a radial turning point and the particle returns to infinity (the in/out two branches of scattered particles). The maximum angular momentum $\bar{L}_{\max}$ allowed for a scattered orbit is determined by the condition that the particle just fails to exist at the current position $(r_{x}, \theta_{x})$, i.e., $(\rho^{2} \dot{r})^{2}=R|_{(r_{x}, \theta_{x})}=0$, which yields
\begin{align}
\bar{L}_{\max}(\sigma, E; r, \theta)&= \frac{1}{\varDelta-(a\sin\theta\sin\sigma)^{2}} \cdot   \left[   -a\sin\theta\sin\sigma(\rho^{2}E-\kappa  r) \right.  \notag \\
& \qquad  \left. +\sqrt{\varDelta \left[(\rho^{2}E-\kappa  r)^{2}-\rho^{2}(\varDelta-(a\sin\theta\sin\sigma)^{2})\right]}\right]. \label{Lmax}
\end{align}
\item
Unphysical trajectory: If $\bar{L} > \bar{L}_{\max}$, the particle cannot reach that point and is excluded.
\end{enumerate}

These constraints formally define the absorption and scattering domains in momentum space:
\begin{align}
D_{\text{abs\; }} &\equiv D_{\sigma}\times D_{E \,\text{abs\,}} \times D_{\bar{L} \,\text{abs\,}} =[0,2\pi] \times [1, +\infty) \times (0, \bar{L}_{c} ), \label{Dabs} \\
D_{\text{scat}}&\equiv D_{\sigma}\times D_{E \,\text{scat}}\times D_{\bar{L} \,\text{scat}}=[0,2\pi] \times [E_{c}, +\infty) \times (\bar{L}_{c},\bar{L}_{\text{max}}). \label{Dscat}
\end{align}
This precise delineation of the phase space domains provides the foundation for computing the momentum-space integrals required for physical observables in the subsequent section.

\section{Physical Observables of the Vlasov Gas System} \label{Sec4}

\subsection{Definition of observable quantities}

Having obtained the absorption domain $D_{\text{abs}}$ and the scattering domain $D_{\text{scat}}$ of particles at any spacetime point $x$, we now define the physical quantities point by point.
The fundamental physical observables characterizing the Vlasov gas system are the particle flux density $J_{\mu}$ and the energy-momentum tensor $T_{\mu \nu}$, defined through momentum-space integrals of the distribution function:
\begin{align}
J_{\mu}|_{x} &\coloneqq \int_{F^{+}_{x}}p_{\mu}f \text{dvol}_{x}(p), & T_{\mu \nu}|_{x} &\coloneqq \int_{F^{+}_{x}}p_{\mu}p_{\nu}f \text{dvol}_{x}(p), \label{JandT}
\end{align}
where $\text{dvol}_{x}(p)$ represents the invariant volume element on the future mass hyperboloid $F^{+}_{x}\equiv \displaystyle\mathop{\cup}_{m>0} F^{+}_{x}(m)$. The physical momentum (as opposed to the canonical momentum) of a particle is
\begin{align}
p_{t}&=-E+\frac{\kappa r }{\rho^{2}}, & p_{r}&=\pm \frac{\sqrt{R}}{\varDelta}, & p_{\theta}&=\bar{L}\cos \sigma, &p_{\varphi} &=a\sin^{2}\theta \left(E- \frac{\kappa  r }{\rho^{2}} \right) +\bar{L}\sin \theta \sin \sigma.  \label{phypmu}
\end{align}
For $p_{r}$, the $+$ sign corresponds to outgoing particles and the $-$ sign to ingoing particles. The invariant volume element can be expressed as:
\begin{align}
\text{dvol}_{x}(p)=\sqrt{|\text{det}[g^{\mu \nu}]|}\text{d}p_{t}\wedge \text{d}p_{r}\wedge \text{d}p_{\theta}\wedge \text{d}p_{\varphi}=\dd{E} \wedge \pqty{ \frac{\dd{P_{0}}}{\sqrt{R/P^{2}_{0}}} } \wedge  (\bar{L} \text{d}\bar{L}) \wedge \text{d}\sigma. \label{dvolxp}
\end{align}
At this point we can write down the integral expressions for $J_{\mu}$ and $T_{\mu \nu}$. The integral expression for $J_{t}$ is
\begin{align}
J^{\text{abs}}_{t}&=-\int_{D_{\text{abs}}}E_{\kappa} \cdot   \frac{ \bar{f}_{4} \cdot  \bar{L}}{\sqrt{R}} \dd{\sigma}\dd{E}\dd{\bar{L}}, & J^{\text{scat}}_{t}&=-2\int_{D_{\text{scat}}}E_{\kappa}  \cdot  \frac{ \bar{f}_{4} \cdot   \bar{L}}{\sqrt{R}} \dd{\sigma}\dd{E}\dd{\bar{L}},  \label{Jtperm}
\end{align}
where $E_{\kappa} \equiv E-\frac{\kappa r}{\rho^{2}}$. Apart from $\bar{f}_{n}$, all quantities are dimensionless, and
\begin{align}
\bar{f}_{n} \equiv \int m^{n}	f(m, m E, m M L_z,  m M L) \dd{m}, \quad m \equiv P_{0},\  E \leftarrow \frac{E}{m}, \ L_{z} \leftarrow \frac{L_{z}}{mM},\ L \leftarrow \frac{L}{mM}. 
\end{align}
In Appendix \ref{componentsofJT}, we give the proof that $J_{\theta}$ and all $T_{\theta \mu}$ with $\mu \neq \theta$ vanish, as well as that $J^{\text{scat}}_{r}$ and $T^{\text{scat}}_{r \mu}$ with $\mu \neq r$ vanish, and we provide the complete integral expressions for the remaining non-zero components. When an index of $J_{\mu}$ or $T_{\mu \nu}$ contains $k$ occurrences of $\theta,\varphi$, the dimensionless integral should be supplemented by a factor $M^{k}$:  each $[p_{\varphi}], [p_{\theta}]$ contributes one factor of $M$.

\subsection{Construction of the distribution function}

We now consider a specific physical scenario: a simple plasma Vlasov gas system consisting of two particle species with charges $+q$ and $-q$ but otherwise identical properties, maintaining overall electrical neutrality. 
	
We assume that all particles originate from a particle source at infinity that obeys the J\"uttner equilibrium distribution \cite{Rioseco-Sarbach(20171)},
\begin{align}
f_{\infty}(E)=\alpha_{0}\delta(P_{0}-m) e^{-z E}, \label{Juttnerfun}
\end{align}
where $\alpha_{0}$ is the normalization coefficient, $z\equiv \frac{mc^{2}}{k_{B}T}$ is the relativistic coldness parameter, and $E$ is measured in units of $mc^{2}, E \leftarrow \frac{E}{mc^{2}}$. The particle flux density $J_{\mu}$ and the energy-momentum tensor $T_{\mu \nu}$ are linear in the distribution function, which allows us to describe the plasma gas as follows:
\begin{align}
f &\equiv \beta f^{(+)}+ (1-\beta) f^{(-)}, & J_{\mu}&=\beta J^{(+)}_{\mu}+ (1-\beta) J^{(-)}_{\mu}, & T_{\mu \nu}&=\beta	T^{(+)}_{\mu \nu}+(1-\beta)T^{(-)}_{\mu \nu}, \label{funcplasma}
\end{align}
where $\beta$ is determined by the condition of electric neutrality at infinity. Since apart from their charge the gas particles we consider are completely identical, we may choose $f^{(\pm)}=f_{\infty}$. As will be seen later, this choice leads to $\beta=\frac{1}{2}$.

In addition to the J\"uttner distribution, a monoenergetic particle source is also an interesting object of study,
\begin{align}
f_{ \epsilon \infty}=\alpha_{\epsilon}\delta(P_{0}-m)\delta(E-\epsilon). \label{deltafun}
\end{align}
Similarly, we may take $f^{(\pm)}=f_{ \epsilon \infty}$, $f=\beta f^{+}+(1-\beta)f^{-}$.

In both models, the mass-weighted moments become, respectively,
\begin{align}
\overline{f^{(\pm)}}_{n} &=\alpha_{0}m^{n}e^{-zE},& \overline{f^{(\pm)}}_{n}&=\alpha_{\epsilon}m^{n}\delta(E-\epsilon).
\end{align}

The assumption of equal mass for both charged species establishes a theoretically clean framework to isolate and highlight the fundamental interplay between spacetime geometry and electromagnetic interactions, free from complications arising from mass disparity.\footnote{In fact, the parameters $\alpha_{0}, m , z$ could differ for different components, and our discussion would still be applicable. However, a more complex multi-fluid model is beyond the scope of our discussion.}

\subsection{Multi-component plasma and frame-dependent observables}

In the plasma model that we are discussing, the total particle flux is given by $J^a = \beta J^{(+)}{}^{a}+ (1-\beta) J^{(-)}{}^{a}$, while the particle number density in the comoving frame is defined as:
\begin{align}
n_{J}\coloneqq \sqrt{-g^{ab}J_{a}J_{b}}. \label{propern}
\end{align}

To extract physically meaningful observables, we employ the Locally Non-Rotating Frame (LNRF) \cite{Bardeen(1972)}, which provides a natural reference frame for measurements by zero-angular-momentum observers. The LNRF tetrad is defined by:
 \begin{subequations}\begin{align}
 (e_{0})^{a}&= e^{-\nu}\left[(\partial_{t})^{a}+\omega (\partial_{\varphi})^{a}\right], &
 (e_{3})^{a}&=e^{-\varPsi}(\partial_{\varphi})^{a},&
 (e_{1})^{a}&=e^{-\mu_{1}}(\partial_{r})^{a}, &
 (e_{2})^{a}&=e^{-\mu_{2}}( \partial_{\theta} )^{a}, \label{LNRFvec}\\
 (e^{3})_{a}&=e^{\varPsi}\left[(\text{d}\varphi)_{a}-\omega (\text{d}t)_{a}\right],&
 (e^{0})_{a}&=e^{\nu}(\text{d}t)_{a},&
 (e^{1})_{a}&=e^{\mu_{1}}(\text{d}r)_{a},&
 (e^{2})_{a}&=e^{\mu_{2}}(\text{d}\theta)_{a},\label{LNRFdual}
\end{align}\end{subequations}
where the metric functions are given by $e^{2\nu}=\frac{\rho^{2}\varDelta}{\varSigma^{2}},\quad e^{2\varPsi}=\frac{\varSigma^{2}\sin^{2}\theta}{\rho^{2}},\quad e^{2\mu_{1}}=\frac{\rho^{2}}{\varDelta},\quad e^{2\mu_{2}}=\rho^{2},\quad \omega =\frac{a(2Mr-Q^{2})}{\varSigma^{2}}.$ We define the simultaneous hypersurface $\Sigma _{t} \equiv \Bqty{ (t, r, \theta, \varphi) \mid t =\text{const}}$, then $e_{0} \perp \Sigma_{t}$, and consequently $\Sigma_{t}$ can be regarded as the ``space'' of the LNRF observer.

In the LNRF tetrad, we use hatted indices $\hat{\mu}, \hat{\nu} \in \Bqty{0, 1, 2, 3}$ to denote tetrad components (the coordinate basis indices are $\mu , \nu \in \Bqty{t, r, \theta, \varphi}$), while the lowercase Latin index $a$ is an abstract index. Any tensor $H^{a}{}_{b}$ can be expressed in the tetrad as
\begin{align}
H^{a}{}_{b}&=H^{\hat{\mu}}{}_{\hat{\nu}}(e_{\hat{\mu}})^{a}(e^{\hat{\nu}})_{b}, &  H^{\hat{\mu}}{}_{\hat{\nu}} & =H^{a}{}_{b} (e^{\hat{\mu}})_{}(e_{\hat{\nu}})^{b},
\end{align}
and the hatted indices are raised and lowered with the metric components $  \eta_{\hat{\mu} \hat{\nu}}=\eta^{\hat{\mu} \hat{\nu}} =\operatorname{diag}[-1,+1,+1,+1] $.

In this frame, the particle number density measured by LNRF observers is:
\begin{align}
n =  -J_{0}=-e^{-\nu}(J_{t}+\omega J_{\varphi}), \label{defn}
\end{align}
while the energy density becomes:
\begin{align}
\varepsilon =T_{00} =e^{-2\nu}(T_{tt}+2\omega T_{t \varphi}+\omega^{2}T_{\varphi \varphi}).
\end{align}
The stress tensor---the spatial projection of the energy-momentum tensor---takes the form:
\begin{align}
\hat{T}^{i}{}_{j}=
\begin{bmatrix}
e^{-2\mu_{1}}T_{rr}&0&e^{-\mu_{1}-\varPsi}T_{r \varphi}\\
0 &e^{-2\mu_{2} }T_{\theta \theta}&0\\
e^{-\mu_{1}-\varPsi}T_{ \varphi r} &0 &e^{-2\varPsi}T_{\varphi \varphi}
\end{bmatrix}.
\end{align}
The principal pressures, corresponding to the eigenvalues of this stress tensor, are given by:
\begin{subequations}\begin{align}
\mathcal{P}_{1}&=\frac{1}{2}\left(\hat{T}^{1}{}_{1}+\hat{T}^{3}{}_{3}+\sqrt{(\hat{T}^{3}{}_{3}-\hat{T}^{1}{}_{1})^{2}+4(\hat{T}^{1}{}_{3})^{2}}\right),\\
\mathcal{P}_{2}&=\hat{T}^{2}{}_{2},\\
\mathcal{P}_{3}&=\frac{1}{2}\left(\hat{T}^{1}{}_{1}+\hat{T}^{3}{}_{3}-\sqrt{(\hat{T}^{3}{}_{3}-\hat{T}^{1}{}_{1})^{2}+4(\hat{T}^{1}{}_{3})^{2}}\right).	
\end{align}\end{subequations}

In order to analyze the difference between the Vlasov gas and a perfect fluid, it is necessary to write down the form of a perfect fluid in the LNRF. Consider a perfect fluid whose four-velocity field is $U^{a}$ with $U^{\theta}=0$ (since $J^{\theta}=0$ for the Vlasov gas). Its energy-momentum tensor is
\begin{align}
T^{\text{fluid}}_{ab}=(\mu+p)U_{a}U_{b}+pg_{ab},
\end{align}
where $U_{a}\coloneqq	g_{ab}U^{b}$, and $\mu, p$ are the proper energy density and proper pressure in the comoving frame, respectively. For an LNRF observer, the stress-tensor components of the perfect fluid are
\begin{align}
[\hat{T}^{\text{fluid}}]^{i}{}_{j}=p\begin{bmatrix}
1 & 0 &0\\
0 &1 &0\\
0 &0 &1
\end{bmatrix}+(\mu+p)\gamma^{2} \begin{bmatrix}
(v^{1})^{2} &0 & v^{1}v^{3}\\
0 &0 &0 \\
v^{1}v^{3} &0 &(v^{3})^{2}
\end{bmatrix},
\end{align}
where $\gamma=-(e^0)_a U^{a}, v^{i}=\gamma^{-1}(e^i)_a U^a$, and the energy density and principal pressures are
\begin{align}
\varepsilon^{\text{fluid}}&=\gamma^{2}(\mu +p)-p, &	\mathcal{P}^{\text{fluid}}_{1}&=\gamma^{2}(\mu+p)-\mu, & \mathcal{P}^{\text{fluid}}_{2}&=p, & \mathcal{P}^{\text{fluid}}_{3}&=p.
\end{align}
From the expressions for $\varepsilon^{\text{fluid}}$ and $\mathcal{P}^{\text{fluid}}_{1}$ we obtain:
\begin{align}
\mu &=\frac{ \varepsilon^{\text{fluid}} -(1- \gamma^{-2}) \mathcal{P}^{\text{fluid}}_{1}}{2-\gamma^{-2}}, & p &= \frac{\mathcal{P}^{\text{fluid}}_{1}-(1-\gamma^{-2}) \varepsilon^{\text{fluid}}}{2-\gamma^{-2}}.
\end{align}

We construct the following diagnostic parameters for the Vlasov gas:
\begin{align}
\lambda_{ \varepsilon} & \coloneqq \frac{1}{n_{J}} \frac{ \varepsilon -(1- \gamma^{-2}) \mathcal{P}_{1}}{2-\gamma^{-2}}, & \lambda_{\text{rad}}& \coloneqq \frac{1}{\mathcal{P}_{2}} \frac{\mathcal{P}_{1}-(1-\gamma^{-2}) \varepsilon}{2-\gamma^{-2}}, & \lambda_{\text{tan}} &\coloneqq \frac{\mathcal{P}_{3}}{\mathcal{P}_{2}},
\end{align}
where $\gamma=\frac{n}{n_J}$.  $\lambda_{ \varepsilon}$ represents the average proper energy per particle, while $\lambda_{\text{rad}}$ and $\lambda_{\text{tan}}$ reflect the anisotropy information of the proper pressures. To some extent, these parameters quantify the non-ideality and anisotropy of the Vlasov gas. When $\lambda_{\text{rad}}\to 1$ and $\lambda_{\text{tan}}\to 1$, the pressure is completely isotropic; when $\lambda_{ \varepsilon}$ simultaneously tends to some constant value and fluctuates very little, the system approaches a perfect fluid.

\subsection{Asymptotic behavior at spatial infinity }

At spatial infinity, the LNRF observers asymptotically become static observers. When analyzing the particle number density $n_{\infty}$, the energy density $\varepsilon_{\infty}$ and the principal pressures $\mathcal{P}_{i\infty}$ of the gas at infinity, we will keep $E_{\kappa} \equiv E-\frac{\kappa r}{\rho^{2}}\simeq E-\frac{\kappa}{r}$ in appropriate places, which helps to analyze in detail the deviation of a charged gas from a neutral gas.

At spatial infinity, the asymptotic expression of the particle flux density vector is (neglecting higher-order small quantities; technical details are given in Appendix \ref{Intinf})
\begin{align}
J^{(\kappa)}_{a}&=-(e^{0})_{a} \cdot 4\pi\int^{\infty}_{1} \bar{f}_{4}(E) E_{\kappa}\sqrt{E^{2}_{\kappa}-1} \dd{E} . \label{J(0)a}
\end{align}
In this case, the Vlasov gas particles are asymptotically at rest, and the proper particle number density coincides with the particle number density defined by the LNRF observer, i.e., $\frac{n_{\infty}}{n_{J\infty}}=1$.

For the J\"uttner distribution, the particle number density of a neutral gas ($\kappa =0$) is
\begin{align}
n^{(0)}_{\infty}&=-J^{(0)}_{0}=(\alpha_{0}m^{4})\cdot4\pi\cdot \frac{K_{2}(z)}{z}, & \alpha_{0}= \frac{n^{(0)}_{\infty}}{4\pi m^{4} \frac{K_{2}(z)}{z}}.
\end{align}
For a charged gas, we expand Eq. \eqref{J(0)a} in $\kappa$ and retain the leading order, obtaining
\begin{align}
J^{(\kappa)}_{a}(\kappa)&\simeq \pqty{1-\frac{z \kappa}{r}}\cdot J^{(0)}_{a} , & n^{(\kappa)}_{\infty}&\simeq \pqty{1-\frac{z \kappa}{r}}\cdot n^{(0)}_{\infty}, \label{nkappan}
\end{align}
which shows that: because of electromagnetic repulsion (assuming the black hole has $Q>0$), a positively charged gas ($\kappa >0$) is slightly rarefied, while a negatively charged gas ($\kappa <0$), due to electromagnetic attraction, is slightly concentrated.

For a plasma gas, the charge density vector should be taken as
\begin{align}
\mathcal{J}_{a}=q \beta J^{(+)}_{a}-q(1-\beta)J^{(-)}_{a}=q \bqty{(2\beta-1)-\frac{z|\kappa|}{r}}J^{(0)}_{a}. \label{qJ}
\end{align}
Neglecting the small quantity $r^{-1}$, the net charge density is $q(2\beta-1)n^{(0)}_{\infty}$, and the electric neutrality of the particle source requires $\beta=\frac{1}{2}$. If one retains the order $r^{-1}$, the net charge density is $-\frac{zq|\kappa|}{r}\cdot n_{\infty}$, and in the process of accretion onto a Kerr-Newman black hole, this residual net charge will slowly neutralize the black hole's charge. We can now obtain the particle flux density vector and the particle number density of the plasma (accurate to order $\kappa$):
\begin{align}
J_{a}&=\frac{1}{2}J^{(+)}_{a}+\frac{1}{2}J^{(-)}_{a}=J^{(0)}_{a}, & n_{\infty}&=n^{(0)}_{\infty}.
\end{align}
	
For all types of gas discussed above, if one neglects $r^{-1}$ and higher-order small quantities at infinity, the particle number densities at infinity are all the same, consistent with the result for a neutral gas in Schwarzschild spacetime \cite{Rioseco-Sarbach(20171)}.

We now analyze the energy-momentum tensor of a charged gas:
\begin{align}
T_{00}&=4\pi \int^{\infty}_{1}\bar{f}_{5}E^{2}_{\kappa}\sqrt{E^{2}_{\kappa}-1}\dd{E},  \\
T_{11}=T_{22}=T_{33} &=\frac{4\pi}{3}\int^{\infty}_{1} \bar{f}_{5}\pqty{E^{2}_{\kappa}-1}^{\frac{3}{2}} \dd{E}. 
\end{align}	
The relative deviation of the energy density $\varepsilon^{(\kappa)}_{\infty}$ and the principal pressures $\mathcal{P}^{(\kappa)}_{i \infty}$ from the neutral case obeys the same relation as the deviation of the particle number density $n^{(\kappa)}_{\infty}$ from $n^{(0)}_{\infty}$ in Eq. \eqref{nkappan}. Accurate to order $\kappa$, the plasma is identical to the neutral gas, i.e.,
\begin{align}
\varepsilon_{\infty}&=m n_{\infty} \left(\frac{K_{1}(z)}{K_{2}(z)}+\frac{3}{z}\right), & \mathcal{P}_{i \infty} &=\frac{m n_{\infty}}{z}, \text{ for } i=1,2,3.
\end{align}

At this stage, we can write down the diagnostic parameters of the Vlasov gas at infinity:
\begin{align}
\lambda_{ \varepsilon}&=m \pqty{ \frac{K_{1}(z)}{K_{2}(z)}+\frac{3}{z}}, &\lambda_{\text{rad}} &=1,  & \lambda_{\text{tan}} &=1.
\end{align}
This is similar in structure to the parameters of a perfect fluid: the proper energy is constant and the pressure is isotropic. The temperature parameter only affects $\lambda_{ \varepsilon}$ and does not alter the isotropic nature of the pressure.

In the low-temperature limit, $K_{n}(z)|_{z \to \infty}\simeq \sqrt{\frac{\pi}{2z}} e^{-z}\pqty{1+\frac{4n^{2}-1}{8z} +o(z^{-1})}$, 
\begin{align}
\varepsilon_{\infty}&=mn_{\infty}\pqty{1+\frac{3}{2z}}= n_{\infty} \pqty{m +\frac{3}{2}k_{B}T}, & \mathcal{P}_{i\infty}&=n_{\infty}k_{B}T, \quad i=1,2,3.
\end{align}
The particle source is approximately an ideal gas, and when $T=0$, the particle source becomes dust. In the high-temperature limit, $K_{n}(z)|_{z \to 0^{+}}=\frac{\Gamma(n)}{2}\pqty{\frac{2}{z}}^{n}$,
\begin{align}
\mathcal{P}_{i\infty}=\frac{1}{3}\varepsilon_{\infty}, \quad i=1,2,3.
\end{align}
which is the equation of state of an ultra-relativistic gas, analogous to a radiation field.

For the monoenergetic distribution (Eq. \eqref{deltafun}), we can immediately write down the physical quantities of a neutral gas:
\begin{align}
n^{(0)}_{\infty}&=4\pi (\alpha_{\epsilon}m^{4})\epsilon \sqrt{\epsilon^{2}-1}, & \alpha_{\epsilon}&=\frac{n^{(0)}_{\infty}}{4\pi m^{4} \epsilon \sqrt{\epsilon^{2}-1}},\\
\varepsilon^{(0)}_{\infty}&=m \epsilon n_{\infty}, & \mathcal{P}^{(0)}_{i\infty}&=\frac{mn_{\infty}}{3}(\epsilon -\epsilon^{-1}), \quad i=1,2,3.
\end{align}
Similarly to the J\"uttner case above, the positive and negative charged gases deviate slightly from the neutral gas:
\begin{align}
n^{(\kappa)}_{\infty} &=n^{(0)}_{\infty}-\frac{\kappa}{r} \dv{\ }{\epsilon}n^{(0)}_{\infty}, &\varepsilon^{(\kappa)}_{\infty} &=\varepsilon^{(0)}_{\infty}-\frac{\kappa}{r} \dv{\ }{\epsilon}\varepsilon^{(0)}_{\infty}, & \mathcal{P}^{(\kappa)}_{i\infty} &=\mathcal{P}^{(0)}_{i\infty}-\frac{\kappa}{r} \dv{\ }{\epsilon}\mathcal{P}^{(0)}_{i\infty},\ i=1,2,3.
\end{align}
The plasma is identical to the neutral gas (accurate to order $\kappa$), and its diagnostic parameters are
\begin{align}
\lambda_{ \varepsilon}&=m \epsilon, & \lambda_{\text{rad}}&=1, & \lambda_{\text{tan}}&=1,
\end{align}
which is precisely a standard perfect fluid.

\subsection{Conservation laws and accretion rates}

The Vlasov system satisfies important conservation laws \cite{Sarbach(2022)}. The particle current is conserved: $\nabla^{a}J_{a} =0$, while the energy-momentum tensor satisfies $\nabla^{a}T_{ab}=\mathcal{F}_{bc}\mathcal{J}^{c}$, where $\mathcal{J}^{a}=\frac{q}{2}\pqty{  J^{(+)}{}^{a}- J^{(-)}{}^{a}}$ is the electric current density, and $\mathcal{F}_{ab}$ is the total electromagnetic field.

Since $J^{(\kappa)}$ depends continuously on the parameter $\kappa$, when $\kappa \ll 1$, we have $\frac{1}{2}\pqty{J^{(+)}-J^{(-)}}\propto \kappa$. On the other hand, we have assumed that all gas particles satisfy the test particle condition ($q \ll Q$), so that the electromagnetic field of the black hole itself, $F_{ab}\ (\propto Q)$, dominates the total electromagnetic field $\mathcal{F}_{ab}$. Consequently, $\mathcal{F}_{bc}\mathcal{J}^{c} \propto Q \cdot q \cdot \kappa \simeq \kappa^{2}$, and further,
\begin{align}
\nabla^{a}T_{ab}=0+o(\kappa)\simeq 0. \label{graT}
\end{align}
Accurate to order $\kappa$, we have $\nabla^{a}T_{ab}\simeq 0$.
This divergence-free property permits the construction of conserved currents associated with the spacetime symmetries:
\begin{align}
\varXi_{a} &\coloneqq -T_{ab}\xi^{b}, & \varPsi_{a} & \coloneqq {T}_{ab}\psi^{b}.
\end{align}

The mass, energy, and angular momentum accretion rates---key quantities characterizing the black hole's evolution---are defined through fluxes across a 2-surface $S$ enclosing the black hole. More precisely, they are given by the flux per unit time\footnote{Using the coordinate time interval $\Delta t=t_{2}-t_{1}$ corresponds to an average with respect to a stationary observer at infinity; using the proper time interval of the LNRF observer $\Delta \tau=\tau(t_{2})-\tau(t_{1})$ can define a local physical quantity. However, due to the null nature of the horizon, the latter local quantity would develop a divergent factor near the horizon.} $\frac{1}{\Delta t}\equiv \frac{1}{t_{2}-t_{1}}$ flowing into the four-dimensional spacetime region $\Omega$ bounded by $\Sigma_{t_{1}}, S\times [t_{1}, t_{2}], \Sigma_{t_{2}}$ (where $\Sigma_{t}\equiv \Bqty{ t=\text{const}}$)\footnote{A more refined boundary construction can avoid the influence of the pathological region of the horizon and its ``interior'' without altering our conclusions.}:
\begin{align}
\frac{1}{\Delta t}\int_{\partial \Omega}J^{a}n_{a} \tilde{ \varepsilon},
\end{align}
where $n^{a}$ is the normal vector on the boundary $\partial \Omega$, $n_{a}=g_{ab}n^{b}$, and $\tilde{ \varepsilon}$ is the induced volume element on the boundary.
Because all our conserved currents are stationary, the fluxes over the past and future boundaries exactly cancel. Gauss's theorem (see, e.g., \cite{Wald(1984)}) guarantees the freedom to choose the surface $S$, and we choose the $r=\text{const}$ surface $S_{r}$. Then $e_{1} \perp S_{r} \times [t_{1}, t_{2}]$, and on this boundary $J^{a}n_{a} \tilde{ \varepsilon}= J^{a}(e^{1})_{a}\ e^{2}\wedge e^{0}\wedge e^{3}=J^{r}\rho^{2}\sin\theta \dd{\theta}\wedge  \dd{t} \wedge\dd{\varphi}$.

Because $J_{\mu}$ and $T_{\mu \nu}$ are linear in the distribution function, the accretion rates of the plasma gas can be expressed as a linear superposition of the positively and negatively charged components (with weight $\beta=\frac{1}{2}$). We write down the mass accretion rate of a ``single-species charged gas'':
\begin{align}
\dot{\mathcal{M}}^{(\kappa)}/m&\coloneqq -\frac{1}{\Delta t}\int J^{r}\rho^{2}\sin\theta \dd{\theta}\dd{t}\dd{\varphi} =2\pi \int^{\pi}_{0}(-J^{r})\rho^{2}\sin\theta \dd{\theta}, \notag\\
&=\pi \int^{\pi}_{0} \dd{\theta}\sin\theta \int^{2\pi}_{0} \dd{\sigma} \int^{\infty}_{1}\dd{E}\bar{f}_{4}\bar{L}^{2}_{c}.\label{DotM} 
\end{align}
The minus sign in the first step ensures that the flux entering the interior of $S_{r}$ is positive; the second step simplifies the expression using the stationarity and axisymmetry of the conserved current; the third step follows from substituting $J^{r}$ and carrying out the algebra. The energy and angular momentum accretion rates are defined as
\begin{align}
	\dot{\mathcal{E}}^{(\kappa)}&\coloneqq \pi \int^{\pi}_{0} \dd{\theta}\sin\theta \int^{2\pi}_{0} \dd{\sigma} \int^{\infty}_{1}\dd{E}\bar{f}_{5} E_{\kappa}\bar{L}^{2}_{c}, \label{DotE} \\
	\dot{\mathcal{L}}^{(\kappa)}&\coloneqq  \pi \int^{\pi}_{0} \dd{\theta}\sin^{2}\theta \int^{2\pi}_{0} \dd{\sigma} \int^{\infty}_{1}\dd{E}\bar{f}_{5}\cdot \pqty{ a \sin\theta E_{\kappa} \bar{L}^{2}_{c}+\sin\sigma \frac{2\bar{L}^{3}_{c}}{3}}. \label{DotL}
\end{align}

Before proceeding further, it is helpful to write down again the explicit expression for $\bar{L}_{c}$, namely Eq. \eqref{EcLc(E)},
\begin{align}
\bar{L}_{c}(\sigma, E; \kappa)=\bar{L}_{c}(a,Q, a\sin\theta, a\sin\sigma, E,\kappa). 
\end{align}

We now analyze the influence of $\kappa$ and the factor $E_{\kappa}=E-\frac{\kappa r}{\rho^{2}}$ on the accretion rates of the plasma gas. We only show the energy accretion rate; the other two are similar.
\begin{align}
\dot{\mathcal{E}}^{\text{plasma}}=\frac{1}{2}\dot{\mathcal{E}}^{(+)}+\frac{1}{2}\dot{\mathcal{E}}^{(-)}.
\end{align}
Since the integration domain is the same, we consider the integrand $E_{\kappa}\bar{L}^{2}_{c}(\sigma, E; \kappa)$. For small $\kappa$, we take its expansion
\begin{align}
E_{\kappa}\bar{L}^{2}_{c}(\sigma, E; \kappa) &=\pqty{ E-\frac{\kappa r}{\rho^{2}}}\pqty{\bar{L}^{2}_{c}(\sigma, E;0)+\kappa \eval{\pdv{\bar{L}^{2}_{c}(\sigma, E; \kappa)}{\kappa}}_{\kappa=0}+o(\kappa)}\notag\\
&=E \bar{L}^{2}_{c}(\sigma, E; 0)-\kappa \pqty{ \frac{r}{\rho^{2}}\bar{L}^{2}_{c}(\sigma, E;0)-E\eval{\pdv{\bar{L}^{2}_{c}(\sigma, E; \kappa)}{\kappa}}_{\kappa=0}}+o(\kappa).
\end{align}
The first term on the right-hand side is the relevant physical quantity for the neutral gas (which effectively means $\kappa=0$); the coefficient of $\kappa$ is a constant independent of $\kappa$. Thus, for the plasma we obtain
\begin{align}
\dot{\mathcal{E}}^{\text{plasma}}=\dot{\mathcal{E}}^{(0)}+o(\kappa),
\end{align}
i.e., in the plasma, the combined effect of the charges of the positive and negative components is $o(\kappa)$, which is self-consistent with the discussion of the energy-momentum tensor at the beginning of this subsection (Eq. \eqref{graT}). The accretion rates of the plasma gas are exactly equal to those of a neutral gas, which means
\begin{align}
\kappa=0,\qquad	E_{\kappa}=E, \qquad \bar{L}_{c}(\sigma, E)=\bar{L}_{c}(a,Q, a\sin\theta_{x}, a\sin\sigma, E,0). 
\end{align}
The structure of the expressions for the neutral gas agrees with the results discussed by Mach et al. in the Kerr spacetime \cite{Mach(20261)}; the difference is that our ``neutral result'' is the result for a plasma gas and that the parameter $\bar{L}_{c}$ is affected by the charge $Q$ of the Kerr-Newman black hole.

We now analyze the variation of the black hole's charge, combining the expression for the charge current density $\mathcal{J}_{a}$ at a large radius $r$, i.e., Eq. \eqref{qJ}. Analogous to the mass accretion rate, we can tentatively discuss the charge accretion rate $\dot{\mathcal{Q}}$ given by an observer at position $r$, from which we obtain the crude estimate
\begin{align}
\dot{\mathcal{Q}}\simeq - \frac{|q|}{m}\cdot \frac{z|\kappa|}{r} \dot{\mathcal{M}},  \qquad \Rightarrow \qquad  Q(t) \cdot M(t)^{\frac{z}{r}\pqty{\frac{q}{m}}^{2}}\simeq C(r). \label{QtMt}
\end{align}
It is termed crude because $\mathcal{J}_{a}$ is not a conserved quantity\footnote{Strictly speaking, the physical charge current $\mathcal{J}_{a}$ is exactly conserved, as it is constructed from the conserved particle currents $J^{\pm}_{a}$  \cite{Sarbach(2022)}.  Eq. \eqref{qJ} represents a truncated expansion, and it is this approximation—not the physical current itself—that fails to be identically conserved.}, and approximate because we already treat the charged particles as test particles and neglect all higher-order terms $o(\kappa)$ in $\kappa$. The above result crudely indicates that the product of the black hole's charge and (a constant power of) its mass is constant; as accretion proceeds, the increase in mass necessarily leads to a decrease in charge.

In the J\"uttner distribution, the accretion rate formulas for the plasma gas ($\kappa \ll 1$, accurate to order $\kappa$, with $\simeq$ denoting the neglect of small quantities $o(\kappa)$), which are identical to those for a neutral gas, are:
\begin{align}
	\frac{\dot{M}}{m n_{\infty}}&\simeq \frac{z}{K_{2}(z)}\cdot \frac{M^{2}}{4}\int^{\pi}_{0} \dd{\theta}\sin\theta \int^{2\pi}_{0} \dd{\sigma} \int^{\infty}_{1}\dd{E}e^{-z E}\bar{L}^{2}_{c},\label{dotMz}\\
	\frac{\dot{\mathcal{E}}}{ \varepsilon_{\infty}}&\simeq\frac{z^{2}}{zK_{1}(z)+3K_{2}(z)}\cdot  \frac{M^{2}}{4}\int^{\pi}_{0} \dd{\theta}\sin\theta \int^{2\pi}_{0} \dd{\sigma} \int^{\infty}_{1}\dd{E}e^{-zE} E \bar{L}^{2}_{c}\label{dotEz},\\
	\frac{\dot{\mathcal{L}}}{ \varepsilon_{\infty}}& \simeq\frac{z^{2}}{zK_{1}(z)+3K_{2}(z)}\cdot  \frac{M^{3}}{4}\cdot \notag\\
	&\qquad\qquad\qquad \int^{\pi }_{0} \dd{\theta}\sin^{2}\theta \int^{2\pi}_{0} \dd{\sigma} \int^{\infty}_{1}\dd{E}e^{-zE}\cdot \pqty{ a \sin\theta E  \bar{L}^{2}_{c}+\sin\sigma \frac{2\bar{L}^{3}_{c}}{3}}.\label{dotLz}
\end{align}
Using the integral formulas in our Appendix, Eqs. \eqref{Bessel}--\eqref{Laplace}, we can obtain the low-temperature and high-temperature limits.
In the low-temperature limit, $z \to \infty$, we have the asymptotic expansion
\begin{align}
\frac{\dot{\mathcal{M}}}{m n_{\infty}}=\frac{\dot{\mathcal{E}}}{ \varepsilon_{\infty}}& \simeq \sqrt{2\pi z}\cdot  \frac{M^{2}}{4\pi}\int^{\pi}_{0} \dd{\theta}\sin\theta \int^{2\pi}_{0} \dd{\sigma}b_{c}(1)^{2} , \\
 \frac{\dot{\mathcal{L}}}{ \varepsilon_{\infty}}&\simeq \sqrt{2\pi z} \cdot \frac{M^{3}}{4\pi}\int^{\pi }_{0} \dd{\theta}\sin^{2}\theta \int^{2\pi}_{0} \dd{\sigma} \pqty{ a \sin\theta   b_{c}(1)^{2}+\sin\sigma \frac{2b_{c}(1)^{3}}{3}},
\end{align}
where we have defined the impact parameter on the critical orbit:
\begin{align}
b_{c}(E)\coloneqq \eval{\frac{\bar{L}_{c}(E)}{E}}_{\kappa=0}=\eval{\frac{\bar{L}_{\text{solve}}(r_{\text{c}}(E))}{E}}_{\kappa=0}.
\end{align}
From our discussion of $E_{\text{solve}}$ and $\bar{L}_{\text{solve}}$, $b_{c}(E)$ is well-defined on $[1, \infty)$ and the limit $b_{c}(\infty)$ exists. For a high-temperature gas, $z \to 0^{+}$, we have
\begin{align}
\frac{\dot{\mathcal{M}}}{m n_{\infty}}=\frac{\dot{\mathcal{E}}}{ \varepsilon_{\infty}}& \simeq   \frac{M^{2}}{4}\int^{\pi}_{0} \dd{\theta}\sin\theta \int^{2\pi}_{0} \dd{\sigma}b_{c}(\infty)^{2} , \\
 \frac{\dot{\mathcal{L}}}{ \varepsilon_{\infty}}&\simeq  \frac{M^{3}}{4}\int^{\pi }_{0} \dd{\theta}\sin^{2}\theta \int^{2\pi}_{0} \dd{\sigma} \pqty{ a \sin\theta   b_{c}(\infty)^{2}+\sin\sigma \frac{2b_{c}(\infty)^{3}}{3}}.
\end{align}
Finally, we discuss the monoenergetic distribution (Eq. \eqref{deltafun}); the accretion rates of the plasma are
\begin{align}
\frac{\dot{\mathcal{M}}}{m n_{\infty}}=\frac{\dot{\mathcal{E}}}{ \varepsilon_{\infty}}& \simeq  \frac{M^{2} \epsilon}{4  \sqrt{\epsilon^{2}-1}}\int^{\pi}_{0} \dd{\theta}\sin\theta \int^{2\pi}_{0} \dd{\sigma}b_{c}(\epsilon)^{2},\\
 \frac{\dot{\mathcal{L}}}{ \varepsilon_{\infty}}&\simeq   \frac{M^{3}\epsilon}{4\sqrt{\epsilon^{2}-1}}\int^{\pi }_{0} \dd{\theta}\sin^{2}\theta \int^{2\pi}_{0} \dd{\sigma} \pqty{ a \sin\theta  b_{c}(\epsilon)^{2}+\sin\sigma \frac{2b_{c}(\epsilon)^{3}}{3}}.
\end{align}
It can be seen from the discussion here that in a certain sense, the low-temperature and high-temperature gases are just the $\epsilon=1$ and $\epsilon=\infty$ monoenergetic gases, respectively.

\section{Numerical Results}\label{Sec5}

This section presents the numerical implementation of the physical quantities discussed in Sec.\,\ref{Sec4}. We first examine the spatial profiles of fundamental gas properties—the particle number densities $n_J$ and $n$, the energy density $\varepsilon$, and the principal pressures $\mathcal{P}_i$—to elucidate the behavior of Vlasov gas in Kerr-Newman spacetime and highlight its distinctive characteristics compared to perfect fluids. Subsequently, we analyze the accretion rates $\dot{\mathcal{M}}$, $\dot{\mathcal{E}}$, and $\dot{\mathcal{L}}$, which quantify the black hole's evolutionary trajectory.

For the analysis of gas properties, we adopt the parameters $M=1$, $a=Q=0.1$, $\kappa = 0, \pm 0.1$, and $z=1$. The accretion rates are investigated across the parameter ranges $a, Q \in [0,1]$. The Vlasov gas and black hole constitute a coupled dynamical system, with the black hole charge $Q$ treated as a variable parameter while maintaining fixed dimensionless charge coupling $\kappa$, corresponding to the weak electromagnetic interaction regime.

\subsection{Spatial profiles of gas properties}

In order to better analyze the gas properties, we regroup and combine $n_{J}, n, \varepsilon, \mathcal{P_{1}}, P_{2}, P_{3}$ into six equivalent physical quantities: the proper particle number density $n_{J}$; the mean single-particle energy measured by the LNRF observer $\frac{ \varepsilon}{n}$, the 3-velocity $v\equiv \sqrt{1-\gamma^{-2}}, \gamma=\frac{n}{n_{J}}$; and the diagnostic parameters for the deviation from a perfect fluid $\lambda_{ \varepsilon}, \lambda_{\text{rad}}, \lambda_{\text{tan}}$.

\begin{figure}[htbp]
\centering
\subfloat[$n_{J}-\ln(r/r_{H})$]{\includegraphics[width=0.48\linewidth]{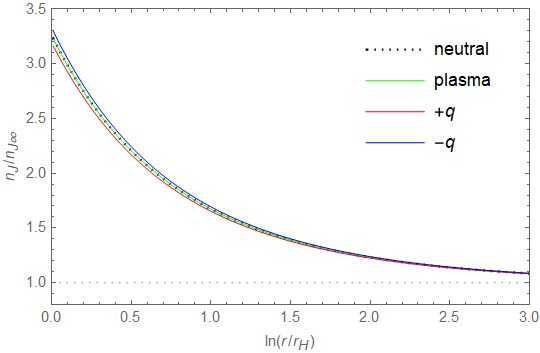} \label{nJrz1}}
\hfill
\subfloat[$n-\ln(r/r_{H})$]{\includegraphics[width=0.48\linewidth]{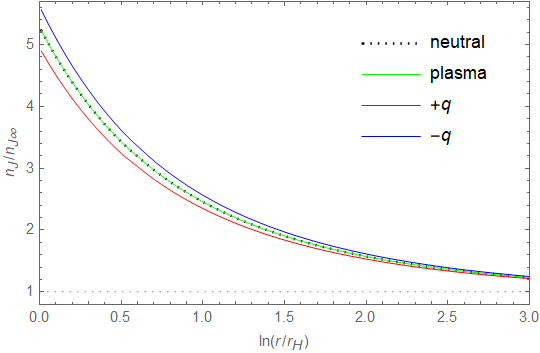} \label{nJrz10}}
\caption{Curves of the proper particle number density as a function of the radial coordinate in the equatorial plane for pure neutral gas, plasma, and pure charged $(\pm q)$ gas in the Kerr-Newman spacetime. Left panel (a): gas with finite temperature $z=1$; right panel (b): low-temperature gas with $z=10$.}
\label{nrz}
\end{figure}

In Fig. \ref{nrz}, we plot the cross-section of the proper particle number density $n_{J}(r, \theta)$ on the equatorial plane $\theta=\frac{\pi}{2}$. Fig. \ref{nJrz1} and Fig. \ref{nJrz10} correspond to gases with $z=1$ and $z=10$, respectively. Every subplot shows that the particle number density curve of the plasma (green solid line) nearly coincides with that of the neutral gas (black dotted line); due to electromagnetic repulsion from the black hole, the positively charged gas (red solid line, $\kappa>0$) is slightly less dense than the neutral gas, while the negatively charged gas (blue solid line, $\kappa<0$), owing to electromagnetic attraction, is slightly denser. The deviation of a single-species charged gas from the neutral gas continuously decreases with increasing distance $r$; at a given location, the deviation is larger for lower temperatures. This finite-position deviation behavior is consistent with the deviation characteristics of our asymptotic results at infinity, Eq. \eqref{nkappan}.

The deviation of the proper particle number density of a single-species charged gas from that of a neutral gas implies, in the plasma case, a non-zero net charge density, with $qn^{+}_{J}-qn^{-}_{J}<0$. As accretion proceeds, these net charges enter the black hole and will neutralize its original charge, which is also consistent with the predictions from the analysis at infinity, Eq. \eqref{qJ} and Eq. \eqref{QtMt}.

\begin{figure}[htbp]
\centering
\subfloat[$\frac{ \varepsilon/n}{ \varepsilon_{\infty}/n_{\infty}}-\ln(r/r_{H}).$]{\includegraphics[width=0.48\linewidth]{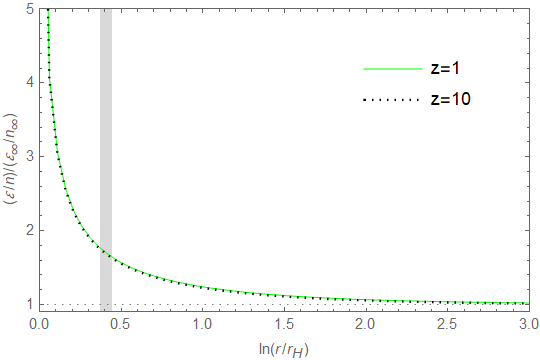} \label{LNRFEbyn}}
\hfill
\subfloat[$v-\ln(r/r_{H})$]{\includegraphics[width=0.48\linewidth]{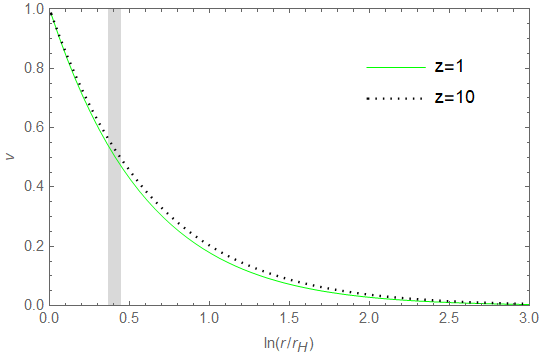} \label{LNRFv}}
\caption{Radial profiles of the mean particle energy $\frac{ \varepsilon}{n}$ (left panel (a)) and the mean 3-velocity $v$ (right panel (b), with $\frac{J^{a}}{n_{J}}$ regarded as the four-velocity field of the gas) in the equatorial plane, as measured by an LNRF observer. The gray rectangular region indicates the possible location of the photon sphere.}
\label{LNRFRbynv}
\end{figure}

Fig. \ref{LNRFRbynv} plots the variation of the mean particle energy and the 3-velocity of the gas particles as measured by an LNRF observer. In Fig. \ref{LNRFEbyn}, a comparison of the curves for $z=1$ and $z=10$ shows that the mean particle energy increases as the distance $r$ from the black hole decreases, and the relative increase is almost independent of the gas temperature. In Fig. \ref{LNRFv}, the velocity of the particles relative to the LNRF observer also increases with decreasing $r$, again nearly independent of temperature. This insensitivity is related to our test-particle assumption: any particle approaching the black hole is subject only to gravitational and electromagnetic interactions. In particular, in the plasma model, the electromagnetic differences experienced by positive and negative charges cancel exactly (to order $\kappa$), so the gas behaves like neutral test particles under gravity alone; consequently, the relative growth rate is almost independent of temperature.

\begin{figure}[htbp]
\centering
\includegraphics[width=0.48\linewidth]{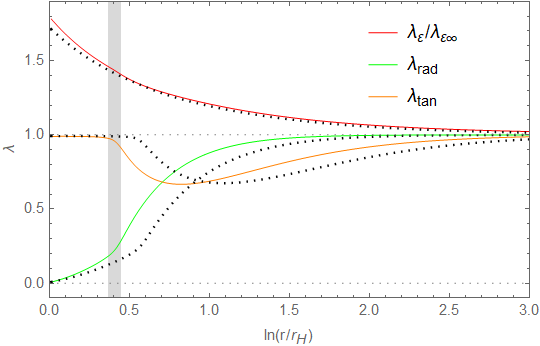} 
\caption{Diagnostic parameters for the deviation of the Vlasov gas from a perfect fluid, constructed in the LNRF frame: the mean proper energy per particle $\lambda_{ \varepsilon}$, the ratio of the radial pressure to the standard tangential pressure ($\mathcal{P}_{2}$), $\lambda_{\text{rad}}$, and the ratio of the second principal tangential pressure to the standard tangential pressure ($\mathcal{P}_{2}$), $\lambda_{\text{tan}}$, as functions of the radial coordinate in the equatorial plane. Colored solid curves correspond to the finite-temperature case $z=1$, while black dotted curves correspond to the low-temperature case $z=10$ (the same diagnostic parameters for the latter case lie very close to the $z=1$ curves, so their labels are omitted).}
\label{EP}
\end{figure}

In Fig. \ref{EP} we show the diagnostic parameter curves of the gas. We obtain the following information on the pressure of the Vlasov gas: far away from the black hole, the pressure is nearly isotropic, $\lambda_{\text{rad}}\simeq \lambda_{\text{tan}}\simeq 1$, and this isotropy is independent of temperature (solid curves for $z=1$, dotted curves for $z=10$). Together with the fact that the mean proper energy $\lambda_{ \varepsilon} \simeq 1$, we see that the Vlasov gas there behaves similarly to a perfect fluid. At finite positions, the Vlasov gas is pressure-anisotropic and differs markedly from a perfect fluid. Close to the horizon (inside the photon sphere), the tangential pressures are nearly equal, while the radial pressure is clearly smaller than the tangential pressures, i.e. $\lambda_{\text{rad}}\ll \lambda_{\text{tan}}\simeq 1$. This discrepancy persists even in Schwarzschild spacetime and may partly explain why the actual accretion rate is far below the Bondi–Michel value \cite{Rioseco-Sarbach(20171)}.

\subsection{Accretion rates and black hole evolution}

We now employ numerical methods to investigate accretion rates for characteristic black hole configurations: Kerr, Reissner-Nordstr\"om(RN), and extremal Kerr-Newman(KN) black holes ($a^{2}+Q^{2} \to 1$), focusing on their dependencies on the rotation parameter $a$ and electric charge $Q$. The accretion rates $\dot{\mathcal{Q}}$ are defined as:
\begin{align}
\dot{\mathcal{Q}}_{ S} &\equiv \dot{\mathcal{Q}}(0,0), & \text{Kerr} &\equiv \dot{\mathcal{Q}}(a,0), & \text{KN}_{a} & \equiv \dot{\mathcal{Q}}\left(a, \sqrt{1-a^{2}}\right), \notag\\
& & \text{RN}& \equiv \dot{\mathcal{Q}}(0,1-Q),& \text{KN}_{Q}& \equiv \dot{\mathcal{Q}}\left(\sqrt{1-Q^{2}},1- Q \right),
\end{align}
with parameters $a, Q \in [0,1]$. In subsequent plots, the abscissa represents $x=a$ for Kerr and Kerr-Newman$_{a}$ curves, while for RN and Kerr-Newman$_{Q}$ curves, it corresponds to $x=1-Q$.

\begin{figure}[htbp]
\centering
\subfloat[$\dot{\mathcal{M}}-x$]{\includegraphics[width=0.48\linewidth]{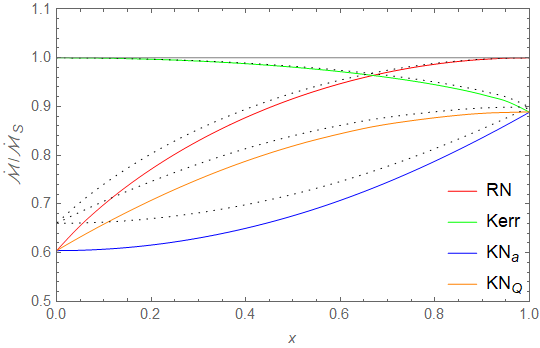}\label{M1} }
\hfill
\subfloat[$\dot{\mathcal{E}}-x$]{\includegraphics[width=0.48\linewidth]{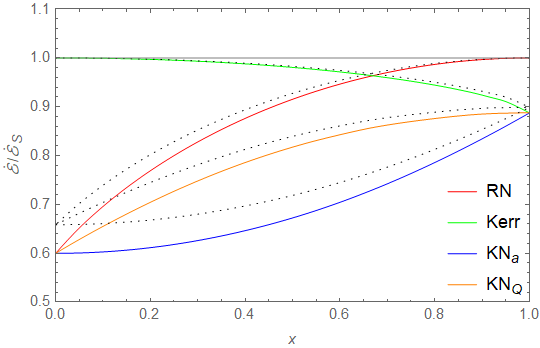} \label{E1}}
\caption{Normalized (a) mass and (b) energy accretion rate parameters for characteristic black holes, scaled by their Schwarzschild counterparts. Solid curves correspond to $z=1$ and dotted curves to $z=10$; since the overall profiles are similar, the labels for the latter are omitted.}
\label{dotM}
\end{figure}

Figs. \ref{M1} and \ref{E1} present the normalized mass and energy accretion rates for characteristic black holes. Fig. \ref{M1} reveals distinct rotational dependencies: the mass accretion rate $\dot{\mathcal{M}}$ for Kerr black holes decreases with increasing rotation parameter $a$, consistent with established theoretical frameworks \cite{Cieslik-Mach(2022)}. Similarly, Reissner-Nordstr\"om black holes exhibit decreasing mass accretion $\dot{\mathcal{M}}$ with increasing charge parameter $Q$. In contrast, extremal Kerr-Newman black holes demonstrate monotonic enhancement of accretion rate parameters with increasing $a$ (equivalently decreasing $Q$). Comparing the cases at different temperatures, we find that the relative accretion rates for the low-temperature case ($z=10$) are overall slightly larger than those for the finite-temperature case ($z=1$), and this deviation becomes gradually more pronounced as $Q$ increases. It is worth noting that, for our chosen plotting parameters, when $z=1$, $\dot{\mathcal{M}}_{S}=89.2, \dot{\mathcal{E}}_{S}=87.2$; when $z=10$, $\dot{\mathcal{M}}_{S}=149.1, \dot{\mathcal{E}}_{S}=144.4$. Therefore, the absolute accretion rates also increase as $z$ increases.

The numerical results reveal identical functional forms between normalized energy and mass accretion parameters $\dot{\mathcal{E}} / \dot{\mathcal{E}}_{ S}$ and $\dot{\mathcal{M}} / \dot{\mathcal{M}}_{ S}$, despite their absolute ratio being $\frac{\dot{\mathcal{E}}}{\dot{\mathcal{M}}}\neq 1$ under our specified parameters. This equivalence originates from the mathematical structure of the integral expressions for $T^{r}{}_{t}$ and $J^{r}$, combined with the exponential decay of the J\"uttner distribution function ($f\propto e^{-zE}$) in particle energy $E \in [1, \infty)$.

\begin{figure}[htbp]
\centering
{\includegraphics[width=0.48\linewidth]{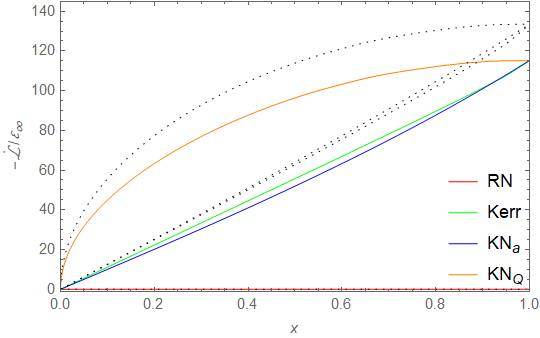}}
\caption{Angular momentum accretion rate parameters for characteristic black holes, normalized by the asymptotic energy density $ \varepsilon_{\infty}$. The distinct behavior compared to mass/energy rates highlights the different physical mechanisms governing angular momentum transfer.}
\label{dotL}
\end{figure}

Fig. \ref{dotL} displays the angular momentum accretion characteristics. As expected, angular momentum accretion vanishes for non-rotating black holes ($a=0$). For Kerr and extremal Kerr-Newman black holes, the angular momentum accretion rate parameters $-\dot{\mathcal{L}}$ increase monotonically with $a$. Unlike the preceding two accretion rates, the charge $Q$ mainly affects the spacetime geometry; it is the rotation parameter $a$ that genuinely determines the angular momentum accretion rate. As with the other two accretion rates, the absolute value of the angular momentum accretion rate, i.e., $-\dot{\mathcal{L}}$, also increases with increasing $z$.

\subsection{Implications for black hole evolution}

The combined analysis of mass accretion rate ($\dot{\mathcal{M}}>0$), particle number density distribution, and angular momentum accretion rate ($-\dot{\mathcal{L}}>0$) yields profound implications for black hole evolution. As accretion proceeds: (1) The black hole mass increases due to positive mass accretion. (2) The net charge $Q$ gradually decreases, driving charged black holes toward electrical neutrality. (3) Electrically neutral black holes maintain their charge neutrality during plasma accretion. (4) The black hole angular momentum decreases gradually due to negative angular momentum accretion.

These evolutionary trends, observed across the investigated parameter space including extreme cases ($a=0$, $Q=0$, $a^2+Q^2 \to 1$), suggest a universal behavior: a general Kerr-Newman black hole surrounded by a plasma Vlasov gas will evolve toward a Schwarzschild configuration through steady-state accretion processes. This result highlights the fundamental role of accretion physics in determining the long-term evolution of astrophysical black holes.

\section{Conclusion}\label{Sec6}

This paper has investigated the kinetic properties and accretion dynamics of collisionless Vlasov gas in Kerr-Newman spacetime. The main conclusions are as follows.

We have demonstrated that spacetime symmetries constrain the distribution function to depend solely on the constants of motion and a configuration variable $Q^{3}$, $f = f(m, E, L_z, L; Q^{3})$. In this paper, we selected the specific J\"uttner distribution $f \propto e^{-zE}$ and the monoenergetic distribution $f \propto \delta(E- \epsilon)$ for analysis.

In the LNRF, we computed physical observables including particle number density, energy density, principal pressures, and accretion rates, and used them to construct a set of diagnostic parameters to describe the deviation of the Vlasov gas from a perfect fluid. For plasma, we derived explicit asymptotic expressions recovering Schwarzschild limits. At infinity, the Vlasov gas behaves like a perfect fluid with isotropic pressure, consistent with our expectations.

Numerical analysis reveals monotonic radial decrease of all physical quantities. While plasma and neutral gas observables closely coincide, charge-dependent asymmetries emerge: positive charges are suppressed, negative charges enhanced. Pressure anisotropy persists even in Schwarzschild spacetime, confirming its statistical origin.

Accretion studies for characteristic black holes show that, for $\kappa \ll 1$, although the accretion rates of a single-species charged gas deviate from those of a neutral gas, the plasma gas agrees with the neutral gas to leading order, with deviations at the order $o(\kappa)\equiv \Bqty{ \kappa^{2}, \kappa^{3}, \cdots}$. Under the J\"uttner distribution, the numerical results indicate:
(1) Identical functional dependence of mass and energy accretion rates due to J\"uttner distribution properties;
(2) Suppression of mass accretion with increasing $a$ and $Q$;
(3) Contrasting angular momentum behavior (for Kerr and extremal Kerr-Newman black holes): increasing with $a$ but decreasing with $Q$;
(4) As the temperature parameter $z$ increases, all accretion rates are enhanced.

These results imply that accretion of weakly charged plasma drives Kerr-Newman black holes toward Schwarzschild configurations through charge neutralization and angular momentum reduction.

\ack{ \ }

\funding{This work was supported by the Young Natural Science Foundation of Wuhan Donghu University under Grant 2025dhzk016.}

\roles{Yongqiang Liu: Conceptualization, Methodology, Formal analysis, Investigation, Software, Validation, Visualization, Writing-original draft, Writing-review \& editing.}

\data{The data that support the findings of this study are included within the article. All mathematical expressions and computational methods are fully described within the manuscript, and the results can be reproduced using the equations provided.}

\appendix

\section{The components of $J_{\mu}$ and $T_{\mu \nu}$}

\subsection{Complete integral expressions}\label{componentsofJT}

We prove that the components $J_{\theta}$ and $T_{\theta \mu}$ with $\mu\neq \theta$ are identically zero. Taking $J^{\text{abs}}_{\theta}$ as an example,
\begin{align}
J^{\text{abs}}_{\theta}=\int^{2\pi}_{0} \bqty{\int^{\infty}_{1}\int^{\bar{L}_{c}(\sigma, E)}_{0} \pqty{\bar{L}\cos\sigma} \frac{\bar{f}_{4} \cdot \bar{L}}{\sqrt{R}}\dd{\bar{L}}\dd{E} }\dd{\sigma}.
\end{align}
From the definitions of $R$ and $\bar{L}_{c}(\sigma, E)$ we see that the inner integrals constitute a function of $\sin \sigma$, denoted $g(\sin\sigma)$. Hence they all belong to the type of zero integral $\int^{2\pi}_{0}\cos\sigma \cdot g(\sin\sigma) \dd{\sigma} =0.$

For the scattered component, the domains of the ingoing and outgoing trajectories are equal, i.e., $D_{\text{scat-in}}=D_{\text{scat-out}}=D_{\text{scat}}$. Since at any spacetime point $x$ we have $p^{\text{scat-in}}_{r}+p^{\text{scat-out}}_{r}=0$, it follows that $J^{\text{scat}}_{r}=J^{\text{scat-in}}_{r}+J^{\text{scat-out}}_{r}=0$. The proof that the components $T^{\text{scat}}_{r \mu}$ with $\mu \neq r$ vanish is analogous.

The non-zero components of the particle flux density vector $J_{\mu}$ are as follows:
\begin{align}
J_{t}&=-\bqty{\int_{D_{\text{abs}}}+ 2\int_{D_{\text{scat}}} } \pqty{ E_{\kappa} \bar{f}_{4}} \cdot \pqty{ \frac{\bar{L}}{\sqrt{R}}}\dd{\bar{L}} \dd{E} \dd{\sigma}, \label{BJt}\\
J_{r}& =- \frac{1}{ \varDelta}\int_{D_{\text{abs}}} \bar{f}_{4} \cdot \bar{L} \dd{\bar{L}} \dd{E} \dd{\sigma},\\
J_{\varphi}&=-a\sin^{2}\theta J_{t}+\sin\theta \bqty{ \int_{D_{\text{abs}}}+2 \int_{D_{\text{scat}}} }  \pqty{ \bar{f}_{4} \sin\sigma}\cdot \pqty{ \frac{\bar{L}^{2}}{\sqrt{R}}} \dd{\bar{L}} \dd{E} \dd{\sigma}.
\end{align}
Note: The absorbed and scattered parts of $J_{\mu}$ can be read off directly from the corresponding identifiers on the right-hand side; the absence of a scattered identifier on the right-hand side of $J_{r}$ indicates that $J^{\text{scat}}_{r}=0$; the expression for $J_{\varphi}$ employs an iterative relation, i.e., it contains $J_{t}$.

The non-zero components of the energy-momentum tensor $T_{\mu \beta}$ are:
\begin{align}
	T_{tt}&=\bqty{ \int_{D_{\text{abs}}} +2 \int_{D_{\text{scat}}} }\pqty{ \bar{f}_{5}  E^{2}_{\kappa} }\cdot \pqty{ \frac{\bar{L}}{\sqrt{R}}}\dd{\bar{L}} \dd{E} \dd{\sigma},   \\
	T_{rr}&=\frac{1}{\varDelta^{2}} \bqty{ \int_{D_{\text{abs}}} +2 \int_{D_{\text{scat}}} }\bar{f}_{5} \cdot \pqty{ \bar{L} \sqrt{R}} \dd{\bar{L}} \dd{E} \dd{\sigma},\\
	T_{\theta \theta}&=\bqty{ \int_{D_{\text{abs}}} +2 \int_{D_{\text{scat}}} } \pqty{ \bar{f}_{5} \cos^{2}\sigma }\cdot \pqty{ \frac{\bar{L}^{3}}{\sqrt{R}}}\dd{\bar{L}} \dd{E} \dd{\sigma},\\
	T_{t r}&= \frac{1}{\varDelta} \int_{D_{\text{abs}}} \pqty{ \bar{f}_{5}  E_{\kappa} }\cdot \bar{L} \dd{\bar{L}} \dd{E} \dd{\sigma} , \\
	T_{t \varphi}&=-a\sin^{2}\theta\,  T_{tt} -\sin\theta \bqty{ \int_{D_{\text{abs}}} +2 \int_{D_{\text{scat}}} } \pqty{\bar{f}_{5} E_{\kappa} \sin\sigma}\cdot \pqty{ \frac{\bar{L}^{2}}{\sqrt{R}}} \dd{\bar{L}} \dd{E} \dd{\sigma} ,\\
	T_{r \varphi}&=-a\sin^{2}\theta \, T_{tr}-\frac{\sin\theta}{\varDelta} \int_{D_{\text{abs}}} \pqty{\bar{f}_{5}\sin\sigma}\cdot \bar{L}^{2}  \dd{\bar{L}} \dd{E} \dd{\sigma} ,\\
	T_{\varphi \varphi}&=-a\sin^{2}\theta \, \pqty{ 2T_{t \varphi}+a\sin^{2}\theta \, T_{tt} }\notag\\
	&\qquad  +\sin^{2}\theta \bqty{ \int_{D_{\text{abs}}} +2 \int_{D_{\text{scat}}} } \pqty{ \bar{f}_{5}\sin^{2}\sigma}\cdot \pqty{ \frac{\bar{L}^{3}}{\sqrt{R}}}  \dd{\bar{L}} \dd{E} \dd{\sigma}. \label{BTphiphi}
\end{align}

The above triple integrals can all be reduced to double integrals. The specific technique is as follows. Let $B$ be a real number, $A, C$ be positive numbers, $G=\sqrt{A^{2}C^{2}+B^{2}}$, and define $F(x) \coloneqq A^{2}-2Bx-C^{2}x^{2}$. Construct the indefinite integrals
\begin{align}
\mathcal{I}_{k} (x) & \coloneqq  \int \frac{ x^{k}}{ \sqrt{F(x)}} \dd{x}, \quad k=0, 1,2,3; &
\mathcal{J}_{k}(x) & \coloneqq  \int x^{k} \sqrt{F(x)} \dd{x}, \quad k=0, 1.
\end{align}
Direct integration yields
\begin{align}
	\mathcal{I}_{0} &= \frac{1}{C} \operatorname{arcsin} \pqty{ \frac{B+C^{2}x}{G} }, &	   \mathcal{I}_{1}  & =-\frac{1}{C^{2}} \pqty{ \sqrt{F} +B \mathcal{I}_{0}},\\
	\mathcal{J}_{0} &=\frac{(B+C^{2}x)\sqrt{F}+G^{2}\mathcal{I}_0}{2C^{2}}, &  \mathcal{J}_{1} &=-\frac{1}{C^{2}}\pqty{ \frac{F^{\frac{3}{2}}}{3} +B \mathcal{J}_{0}},\\
	\mathcal{I}_{2} &=-\frac{1}{C^{2}} \pqty{ \mathcal{J}_{0}+2B \mathcal{I}_{1}-A^{2}\mathcal{I}_{0} }, &  
\mathcal{I}_{3}  &=-\frac{1}{C^{2}}\pqty{ \mathcal{J}_{1}+2B \mathcal{I}_{2}-A^{2}\mathcal{I}_{1} }.
\end{align}
In our case, we set $F(\bar{L})=R$, i.e.,
\begin{align}
F(\bar{L}) &=\left(\rho^{2}E-a\sin\theta\sin\sigma \bar{L}  -\kappa r\right)^{2}-\varDelta (\bar{L}^{2}+\rho^{2})\notag \\
&=\bqty{(\rho^{2}E_{\kappa})^{2}-\varDelta \rho^{2}}-2\bqty{ a \rho^{2} E_{\kappa}\sin\theta\sin\sigma}\bar{L}-\bqty{ \varDelta -(a\sin\theta\sin\sigma)^{2}}\bar{L}^{2},\\
A^{2}&= (\rho^{2}E_{\kappa})^{2}-\varDelta \rho^{2}, \qquad B= a \rho^{2} E_{\kappa}\sin\theta\sin\sigma, \qquad   C^2 =\varDelta -(a\sin\theta\sin\sigma)^{2},\\
G^{2}&=\varDelta \bqty{ A^{2}+\rho^{2}(a\sin\theta\sin\sigma)^{2}}.
\end{align}
Outside the horizon we always have $A^{2}>0$, and outside the static limit ($\varDelta -(a\sin\theta)^{2}>0$) we always have $C^{2}>0$. In the ergoregion between the horizon and the static limit, if $C^{2}<0$ occurs, we set $C\equiv \mathrm{i} \sqrt{|C^{2}|}$ and use the complex identities of inverse trigonometric functions to convert the integral to a real one; the special case $C^{2}=0$ is handled by taking a limit.

Using these primitives, the triple integrals over $(\sigma, E, \bar L)$ in Eqs.\eqref{BJt}--\eqref{BTphiphi} can be reduced to double integrals over $(\sigma, E)$ by evaluating the $\bar L$--integration exactly:
\begin{align}
\int \bar L\sqrt{R}\,\dd\bar L  &= \mathcal{J}_1(\bar L_{\rm top}) - \mathcal{J}_1(\bar L_{\rm bot}),   &
\int \frac{\bar L}{\sqrt{R}}\,\dd\bar L   &= \mathcal{I}_1(\bar L_{\rm top}) - \mathcal{I}_1(\bar L_{\rm bot}), \notag\\
\int \frac{\bar L^2}{\sqrt{R}}\,\dd\bar L  &= \mathcal{I}_2(\bar L_{\rm top}) - \mathcal{I}_2(\bar L_{\rm bot}), &
\int \frac{\bar L^3}{\sqrt{R}}\,\dd\bar L  &= \mathcal{I}_3(\bar L_{\rm top}) - \mathcal{I}_3(\bar L_{\rm bot}),
\end{align}
with the integration limits
$(\bar L_{\rm bot},\bar L_{\rm top}) = (0,\bar L_c)$ 
for absorbed particles and  $(\bar L_{\rm bot},\bar L_{\rm top}) = (\bar L_c,\bar L_{\max})$ 
for scattered ones. At spatial infinity, the leading asymptotic behaviour of $\mathcal{I}_k$ and $\mathcal{J}_k$ at the three boundaries $0$, $\bar L_c$, and $\bar L_{\max}$ is collected in Table~\ref{tab:asympt}, which provides the basis for the asymptotic analysis carried out in the next subsection.

\subsection{Asymptotic expansion of the integrals}\label{Intinf}

When we analyze the asymptotic behavior of the integrals at infinity, we need the values of the functions $\mathcal{I}_{k}(x)$ and $\mathcal{J}_{k}(x)$ at $x=0, \bar{L}_{c}, \bar{L}_{\max}$. Their leading orders are shown in Table \ref{tab:asympt}.
\begin{table}[htbp]
\centering
  \renewcommand{\arraystretch}{1.8}
\begin{tabular}{|c|l|l|l|}
\hline
\textbf{Function} & \textbf{\(x=0\)} & \textbf{\(x=\bar L_c\)} & \textbf{\(x=\bar L_{\max}\)} \\
\hline
\(\mathcal{J}_0(x)\) &
  \(a_{s} E_{\kappa}   r^{2}\sqrt{ E_{\kappa} ^{2}-1} + o(r^{2})\) &
  \(\bigl(a_{s} E_{\kappa}  +\bar L_c\bigr)r^{2}\sqrt{ E_{\kappa} ^{2}-1} + o(r^{2})\) &
  \(\frac{\pi}{4} r^{3}( E_{\kappa} ^{2}-1) + o(r^{3})\) \\[4pt]
\hline
\(\mathcal{J}_1(x)\) &
  \(-\frac{( E_{\kappa} ^{2}-1)^{3/2}}{3} r^{4} + o(r^{4})\) &
  \(-\frac{( E_{\kappa} ^{2}-1)^{3/2}}{3} r^{4} + o(r^{4})\) &
  \(-\frac{\pi a_{s} E_{\kappa}  }{4} r^{3}( E_{\kappa} ^{2}-1) + o(r^{3})\) \\[4pt]
\hline
\(\mathcal{I}_0(x)\) &
  \(\frac{a_{s} E_{\kappa}  }{r^{2}\sqrt{ E_{\kappa} ^{2}-1}} + o(r^{-2})\) &
  \(\frac{a_{s} E_{\kappa}  +\bar L_c}{r^{2}\sqrt{ E_{\kappa} ^{2}-1}} + o(r^{-2})\) &
  \(\frac{\pi}{2r} + o(r^{-1})\) \\[4pt]
\hline
\(\mathcal{I}_1(x)\) &
  \(-\sqrt{ E_{\kappa} ^{2}-1} + o(1)\) &
  \(-\sqrt{ E_{\kappa} ^{2}-1} + o(1)\) &
  \(-\frac{\pi a_{s} E_{\kappa}  }{2r} + o(r^{-1})\) \\[4pt]
\hline
\(\mathcal{I}_2(x)\) &
  \(2a_{s} E_{\kappa}  \sqrt{ E_{\kappa} ^{2}-1} + o(1)\) &
  \(2a_{s} E_{\kappa}  \sqrt{ E_{\kappa} ^{2}-1} + o(1)\) &
  \(\frac{\pi}{4} r ( E_{\kappa} ^{2}-1) + o(r)\) \\[4pt]
\hline
\(\mathcal{I}_3(x)\) &
  \(-\frac{2}{3}( E_{\kappa} ^{2}-1)^{3/2} r^{2} + o(r^{2})\) &
  \(-\frac{2}{3}( E_{\kappa} ^{2}-1)^{3/2} r^{2} + o(r^{2})\) &
  \(-\frac{3\pi a_{s} E_{\kappa}  }{4} r( E_{\kappa} ^{2}-1) + o(r)\) \\[4pt]
\hline
\end{tabular}
\caption{Leading asymptotic behaviour of the integrals \(\mathcal{I}_k\) and \(\mathcal{J}_k\) at the integration boundaries. We denote \( a_{s}\equiv a\sin\theta\sin\sigma\). In addition, retaining \(E_{\kappa}\simeq E -\frac{\kappa}{r}\) facilitates the analysis of the slight deviation of a charged gas from a neutral gas.}
\label{tab:asympt}
\end{table}

The asymptotic values of $J_{\hat{\mu}}$ and $T_{\hat{\mu}\hat{\nu}}$ appearing in the main text at spatial infinity are (apart from $E_{\kappa}$, neglecting all other small quantities $o(1)\equiv \Bqty{r^{-1}, r^{-2}, \cdots}$)
\begin{align}
J_{0}&=-4\pi \int^{\infty}_{1}\bar{f}_{4} \cdot E_{\kappa}\sqrt{E^{2}_{\kappa}-1} \dd{E},\label{BJhat0}\\
T_{00}&=4\pi \int^{\infty}_{1}\bar{f}_{5}E^{2}_{\kappa}\sqrt{E^{2}_{\kappa}-1}\dd{E}, \label{BThat00} \\
T_{11}=T_{22}=T_{33} &=\frac{4\pi}{3}\int^{\infty}_{1} \bar{f}_{5}\pqty{E^{2}_{\kappa}-1}^{\frac{3}{2}} \dd{E}. \label{BThatkk}
\end{align}

When $\kappa=0$, this corresponds to an electrically neutral gas. For the monoenergetic distribution, $\bar{f}_{n} \propto \delta(E-\epsilon)$, we have (omitting the coefficient $\alpha_{\epsilon} m^{n}$ in $\bar{f}_{n}$)
\begin{align}
J_{0}&=-4\pi \epsilon \sqrt{\epsilon^{2} -1},& T_{00}&=4\pi	\epsilon\sqrt{\epsilon^{2}-1}, & T_{11}=T_{22}=T_{33} &=\frac{4\pi}{3}\pqty{\epsilon^{2}-1}^{\frac{3}{2}}.
\end{align}
For the J\"uttner distribution, $\bar{f}_{n} \propto e^{-z E}$, we have (omitting the coefficient $\alpha_{0} m^{n}$ in $\bar{f}_{n}$)
\begin{align}
J_{0}&=-4\pi  \frac{K_{2}(z)}{z} ,& T_{00}&=4\pi \pqty{ \frac{K_{1}(z)}{z} +\frac{3K_{2}(z)}{z^{2}}}, & T_{11}=T_{22}=T_{33} &=4\pi \frac{K_{2}(z)}{z^{2}},
\end{align}
where we have used the special function integral
\begin{align}
\int^{\infty}_{1}E(E^{2}-1)^{\alpha}e^{-z E}\mathrm{d}E =\frac{\Gamma(1+\alpha)}{\sqrt{\pi}}\left(\frac{2}{z}\right)^{\frac{1}{2}+\alpha}K_{\frac{3}{2}+\alpha}(z),
\end{align}
$\Gamma(x)$ is the Euler gamma function, and $K_{\nu}(z)$ is the solution of the second kind Bessel equation $\frac{\mathrm{d}^{2}y}{\mathrm{d}z^{2}}+\frac{1}{z}\frac{\mathrm{d}y}{\mathrm{d}z}-\left(1+\frac{\nu^{2}}{z^{2}}\right)y=0$.

Moreover, for positive integers $n$ and positive real $z$, we have the asymptotic expressions for $K_{n}(z)$
\begin{align}
K_{n}(z) |_{z \to \infty} & \simeq \sqrt{\frac{\pi}{2 z}}e^{-z}\pqty{ 1+ \frac{4n^{2}-1}{8z}+o(z^{-1}) },& K_{n}(z) |_{z \to 0} & \simeq \frac{\Gamma(n)}{2} \pqty{ \frac{2}{z} }^{n}. \label{Bessel}
\end{align}
On the other hand, let $h(x)$ be smooth at $x=1$ and possess a limit as $x\to \infty$. Then the following asymptotic formula for Laplace-type integrals holds:
\begin{align}
\int^{\infty}_{1}e^{-z x}x^{k}h(x) \dd{x} \sim \begin{cases}
\frac{e^{-z}}{z}h(1), & z \to \infty,\\
\frac{\Gamma(k+1)}{z^{k+1}}h(\infty), & z \to 0^{+} , k >-1.
\end{cases} \label{Laplace}
\end{align}
These asymptotic expansions facilitate our analysis of low-temperature gas ($z \to \infty$) and high-temperature gas ($z \to 0^{+}$).

When $\kappa \neq 0$, combining the previous results, i.e., that Eqs. \eqref{BJhat0}--\eqref{BThatkk} depend continuously on $\kappa$, the structure of these integrals is
\begin{align}
\int^{\infty}_{1}\bar{f}_{n} g(E_{\kappa}) \dd{E}, \qquad g(1)=0.
\end{align}
As $\kappa \to 0$, we have $g(E_{\kappa})\simeq g(E)-\frac{\kappa}{r}g'(E)+o(\kappa)$. Hence,
\begin{align}
\int^{\infty}_{1}e^{-z E} g(E_{\kappa}) \dd{E}&\simeq \pqty{ 1- \frac{z \kappa}{r}} \int^{\infty}_{1}e^{-z E} g(E) \dd{E}+o(\kappa),\\
\int^{\infty}_{1}\delta(E-\epsilon) g(E_{\kappa}) \dd{E}&\simeq g(\epsilon)-\frac{\kappa}{r}\dv{g(\epsilon)}{\epsilon}+o(\kappa).
\end{align}

\bibliographystyle{iopart-num}

\bibliography{References}

\end{document}